\documentclass[12pt]{article}%
\usepackage{amssymb}
\usepackage{amsmath}
\usepackage{amsfonts}
\usepackage{graphicx}%
\providecommand{\U}[1]{\protect\rule{.1in}{.1in}}
\begin{document}

\title{\textbf{Scalar torsion and a new symmetry of general relativity}}
\author{J. B. Fonseca-Neto,\thanks{jfonseca@fisica.ufpb.br} C. Romero
\thanks{cromero@fisica.ufpb.br} and S. P.G.
Martinez\thanks{sgomezm@fisica.ufpb.br}}
\maketitle

\begin{abstract}
We reformulate the general theory of relativity in the language of
Riemann-Cartan geometry. We start from the assumption that the space-time can
be described as a non-Riemannian manifold, which, in addition to the metric
field, is endowed with torsion. In this new framework, the gravitational field
is represented not only by the metric, but also by the torsion, which is
completely determined by a geometric scalar field. We show that in this
formulation general relativity has a new kind of invariance, whose invariance
group consists of a set of conformal and gauge transformations, called Cartan
transformations. These involve both the metric tensor and the torsion vector
field, and are similar to the well known Weyl gauge transformations. By making
use of the concept of Cartan gauges, we show that, under Cartan
transformations, the new formalism leads to different pictures of the same
gravitational phenomena. We show that in an arbitrary Cartan gauge general
relativity has the form of a scalar-tensor theory. In this approach, the
Riemann-Cartan geometry appears as the natural geometrical setting of the
general relativity theory when the latter is viewed in an arbitrary Cartan
gauge. We illustrate this fact by looking at the one of the classical tests of
general relativity theory, namely the gravitational spectral shift. Finally,
we extend the concept of space-time symmetry to the more general case of
Riemann-Cartan space-times endowed with scalar torsion. As an example, we
obtain the conservation laws for auto-parallel motions in a static spherically
symmetric vacuum space-time in a Cartan gauge, whose orbits are identical to
Schwarzschild orbits in general relativity. {PACS numbers: 04.20.-q, 04.20.Cv,
04.50.Kd}

\end{abstract}

keywords: {torsion; general relativity; conformal transformations.}

address: {Departamento de F\'{\i}sica, Universidade Federal da Para\'{\i}ba,
Jo\~{a}o Pessoa, PB 58059-970, Brazil}

\section{Introduction}

As we know, Einstein's general relativity is a geometric theory of gravitation
in which the gravitational field appears as a manifestation of the space-time
curvature. According to general relativity, space-time is represented by a
(3+1)-dimensional differentiable manifold endowed with a Lorentzian metric and
a Levi-Civita connection, the latter being completely determined by the
former. However, it is possible to consider a more general geometrical
setting, in which the metric and the connection may be taken as independent
structures. A general metric-compatible connection has two important
properties: curvature and torsion \cite{Hehl,Schouten}. Torsion is \textit{a
priori} absent in the geometrical framework of general relativity, although it
was taken into consideration by Einstein in his search for a unified theory of
gravitation and electromagnetism \cite{Goenner}. Since then, the meaning and
role of torsion in gravitation has been a recurring theme. Initially, Cartan
suggested that torsion could be related to the intrinsic spin of elementary
particles. This idea has developed into the Einstein-Cartan theory, where not
only mass, but also the intrinsic spin of matter were considered as source of
the gravitational field \cite{Cartan}. Further generalization led to
Poincar\"{\i}\textquestiondown
${\frac12}$
gauge theories \cite{Hehl}, in which Lagrangians with quadratic terms in the
curvature and in the torsion have been considered. In all these theories,
torsion is regarded as an independent geometrical structure, which is related
to the intrinsic spin of elementary particles and expected to predict new
gravitational effects. In connection with this, a teleparallel theory of
gravitation was developed, where the presence of the gravitational field
appears as a manifestation of the torsion in a space-time with no
Riemann-Cartan curvature. This approach is now known as the teleparallel
equivalent of general relativity (TEGR), where only mass is assumed to be the
source of torsion, which, in turn, is no longer related to the intrinsic spin
of elementary particles \cite{Aldrovandi}. Finally, let us mention the recent
f(T) theories of gravitation \cite{Aldrovandi}, which are recent
generalizations of TEGR analogous to the recent f(R) theories, mainly
developed to address open questions, such as, the origin of dark matter and
dark energy \cite{fr}. These efforts have given rise to the so-called extended
theories of gravity \cite{Capozziello1} .

In this work, we are concerned with the meaning and role of torsion in gravity
theories from a different point of view by considering two questions of a
rather general character. The first question is: Is it possible to enlarge the
symmetries of general relativity in order to include conformal transformations
of the metric of space-time? The answer to this question seems to have some
relevance as it has been argued recently that a conformal theory of gravity
would be a viable candidate for a quantum theory of gravity, since it is
expected to be renormalizable \cite{Mannheim}. The second questions is: Is it
possible to formulate general relativity in the framework of a Riemann-Cartan
space-time? This is related to the more general question: To what extent is
Riemannian geometry the only possible geometrical setting for general
relativity? In this paper, we will try to address, at least partially, these
two questions.

As is well known, Einstein's theory of gravity in its original formulation is
not invariant under conformal transformations. One reason for this is that the
geometrical language of Einstein`s theory is completely based on Riemannian
geometry. Indeed, for a long time general relativity has been inextricably
associated with the geometry of Riemann. Further developments, however, have
led to the discovery of different geometrical structures, which we might
generically call ``non-Riemannian'' geometries. Many of these developments
were closely related to attempts at unifying gravity with electromagnetism
\cite{Goenner}. Our main objective in this paper is to show that one can
formulate general relativity using the language of a non-Riemannian geometry,
namely, the Riemann-Cartan geometry with a vectorial torsion. In this
formulation, general relativity appears as a theory in which the gravitational
field is described simultaneously by two geometrical fields: the metric tensor
and the torsion field, the latter being an essential part of the geometry,
manifesting its presence in almost all geometrical phenomena, such as
curvature, geodesic motion, and so on. As we will see, in this new geometrical
setting general relativity exhibits a new kind of invariance, namely, the
invariance under Cartan transformations.

The outline of this paper is as follows. We begin by presenting, in Section 2,
the basic mathematical facts of Riemann-Cartan geometry and the concept of
Cartan gauge. In Section 3, we show how to formulate general relativity in a
Riemann-Cartan space-time endowed with scalar torsion. In this formulation, we
will see that the theory is invariant under the group of Cartan
transformations. We devote Section 4 to examine the Newtonian limit, which
will help us to get some insight into the role of the scalar torsion field in
the Cartan representation of general relativity. Then, in Section 5 we
illustrate how different pictures of the same phenomena may arise in distinct
gauges by considering the description of the gravitational spectral shift in a
Cartan gauge. In Section 6, we extend the concept of space-time symmetry to
the more general case of Riemann-Cartan space-times with scalar torsion. As an
example, we obtain the conservation laws for auto-parallel motion and
determine the orbits of test particles, in a static spherically symmetric
vacuum space-time in a Cartan gauge, in the equivalence class of Schwarzschild
space-time. Section 7 contains our final remarks.

\section{Riemann-Cartan space-time and scalar torsion}

In this section, we briefly review some basic properties of Riemann-Cartan
space-times. We will also introduce the notations and conventions that will be
used in this paper. As is well known, in a general affine geometry the metric
and the connection are independent structures. The metric defines the concepts
of length and angle between two vectors, while the connection leads to the
concept of parallel transport and covariant derivative. In such a geometry,
three tensors are of special interest. The first two are the torsion tensor
and the curvature tensor, both depending only on the connection. The
components of the torsion tensor are given, in a coordinate basis, by the
antisymmetric part of the connection coefficients $\mathcal{T}_{\;\;\;\mu\nu
}^{\alpha}=\Gamma_{\;\:\mu\nu}^{\alpha}-\Gamma_{\;\:\nu\mu}^{\alpha}$. \ On
the other hand, the components of the curvature tensor are given by
$R_{\;\:\beta\mu\nu}^{\alpha}=\Gamma_{\;\:\mu\beta,\nu}^{\alpha}-
\Gamma_{\;\:\nu\beta,\mu}^{\alpha}+\Gamma_{\;\:\nu\rho}^{\alpha}%
\Gamma_{\;\:\mu\beta}^{\rho} -\Gamma_{\;\:\mu\rho}^{\alpha}\Gamma
_{\;\:\nu\beta}^{\rho}$. The covariant derivative of a contravariant vector
field $u^{\mu}$, in a coordinate basis, will be given by $\nabla_{\alpha
}u^{\mu}=u_{\;\;;\alpha}^{\mu}=u_{\;\;,\alpha}^{\mu}+\Gamma_{\;\:\alpha\nu
}^{\mu}u^{\nu}$, where $u_{\;\;,\alpha}^{\mu}$ denotes partial derivatives. In
the presence of curvature and torsion, the second covariant derivatives do not
commute and satisfy the Ricci identities $(\nabla_{\beta}\nabla_{\alpha}-
\nabla_{\alpha}\nabla_{\beta})u^{\mu}=R_{\;\;\lambda\alpha\beta}^{\mu
}u^{\lambda}+ \mathcal{T}_{\;\;\;\alpha\beta}^{\lambda}\nabla_{\lambda}u^{\mu
}$ and $(\nabla_{\beta}\nabla_{\alpha}-\nabla_{\alpha}\nabla_{\beta})f=
\mathcal{T}_{\;\;\;\alpha\beta}^{\lambda}\nabla_{\lambda}f\:,$ where $f$ is a
scalar field. Finally, the third important tensor is the non-metricity tensor
defined by the covariant derivative of the metric $Q_{\alpha\mu\nu}%
=\nabla_{\alpha}g_{\mu\nu}$, which depends on both the metric and the
connection (see, for instance, \cite{Hehl,Schouten} and references therein).

Let us recall that the space-time of general relativity corresponds to the
special case in which both the torsion and non-metricity tensors vanish. The
first condition implies that, in a coordinate basis, the connection is
symmetric, while the second condition assures that the connection is
compatible with the metric, that is, lengths and angles between vectors are
assumed to be preserved under parallel transport. We are thus led to the
Levi-Civita connection $\widetilde{\Gamma}$, whose components are known as the
Christoffel symbols of second kind and are given by $\widetilde{\Gamma
}_{\;\:\mu\nu}^{\alpha}=\{_{\mu\nu}^{\alpha}\}= \frac{1}{2}g^{\alpha\beta
}(g_{\beta\nu,\mu}+g_{\mu\beta,\nu}-g_{\mu\nu,\beta})$ \footnote{From now on,
we use tilde superscripts to denote any quantity that depends only on the
metric $g_{\mu\nu}$ and/or the Christoffel symbols calculated with $g_{\mu\nu
}$. The metric assignature is $(+---)$.} . We, thus, see that in the
Riemannian space-time of general relativity the curvature tensors and all
geometric properties of the space-time constructed with $\widetilde{\Gamma}$
will depend only on the metric.

It turned out that shortly after the formulation of general relativity, the
desire to develop a unified and geometrized theory of gravitation and
electromagnetism led physicists to weaken the two restrictive conditions
mentioned above. This would allow to increase the number of degrees of freedom
coming exclusively from geometry. In this way, the first non-Riemannian
manifolds were developed (for details see \cite{Goenner} and references
therein). Let us mention two of them. The first is Weyl geometry, where
torsion is absent, while non-metricity is present. In this case, the
non-metricity tensor is given by $Q_{\alpha\mu\nu}=\omega_{\alpha}g_{\mu\nu}$,
where $\omega_{\alpha}$ is a covariant vector defined on the manifold
\cite{Weyl}. Weyl identified the vector $\omega_{\alpha}$ with the
electromagnetic potential, while the usual gauge transformations of
electromagnetism appear in a natural way as part of the so-called Weyl
transformations. Although his unified theory is no longer considered viable
from the standpoint of physics, Weyl geometry has been revived in a particular
version, known as the Weyl integrable geometry, when $\omega_{\alpha}$
corresponds to the gradient of a scalar field \cite{Scholz}. \ Weyl integrable
geometry has received a lot of attention in the recent years as the
mathematical framework of some alternative gravity theories \cite{Novello}.
The second example of a non-Riemannian geometrical structure is due to H.
Cartan and is called Riemann-Cartan geometry \cite{Cartan}. Here, the
space-time manifold is allowed to possess torsion, while the connection is
required to be compatible with the metric, i.e. $Q_{\alpha\mu\nu}=0$, the
connection has only one restriction: it must be compatible with the metric. In
this case, the metric compatibility condition allows the Riemann-Cartan
connection to be written as $\Gamma_{\;\:\mu\nu}^{\alpha}=\{_{\mu\nu
}^{\:\alpha}\}+K_{\;\:\mu\nu}^{\alpha}$, where $K_{\;\:\mu\nu}^{\alpha}%
=\frac{1}{2}g^{\alpha\beta}(g_{\rho\mu}\mathcal{T}_{\;\;\beta\nu}^{\rho}
+g_{\rho\nu}\mathcal{T}_{\;\;\beta\mu}^{\rho}+g_{\rho\beta}\mathcal{T}%
_{\;\;\mu\nu}^{\rho})$ is the contortion tensor. Let us remark at this point
that in a Riemann-Cartan manifold there are two equivalent sets of independent
fundamental geometric objects, namely, $\{g_{\mu\nu},\Gamma_{\;\:\mu\nu
}^{\alpha}\}$ and $\{g_{\mu\nu},T_{\;\:\mu\nu}^{\alpha}\}$, since the
connection is completely determined by both the metric and the torsion.

An interesting fact is that, in a Riemann-Cartan space-time, there are
transformations involving both the metric and the connection, which preserve
the metric compatibility condition. In addition, these transformations leave
invariant the curvature and, at the same time, change the torsion in a way
similar to a gauge transformation \cite{Shapiro}. They are defined by the
combined effect of a conformal transformation of the metric
\begin{equation}
\overline{g}_{\mu\nu}=e^{f}g_{\mu\nu}, \label{eq:conf trans}%
\end{equation}
where $f$ is a function of the coordinates, and the so-called Einstein's
$\lambda$-transformations of the components of the connection
\cite{Lambd_transf}
\begin{equation}
\overline{\Gamma}_{\;\:\mu\nu}^{\,\alpha}=\Gamma_{\;\:\mu\nu}^{\alpha}%
+\frac{1}{2}\, f_{,\mu}\delta_{\;\:\nu}^{\alpha}. \label{eq:conn trans}%
\end{equation}
Note that\ (\ref{eq:conf trans}) and (\ref{eq:conn trans})\ do not involve
coordinate transformations. It is easy to verify that under these
transformations, the components of the torsion transform according to
\begin{equation}
\overline{\mathcal{T}}_{\;\;\mu\nu}^{\:\alpha}=\mathcal{T}_{\;\;\mu\nu
}^{\alpha} -\frac{1}{2}(\delta_{\;\;\mu}^{\alpha}f_{,\nu}-\delta_{\;\;\nu
}^{\alpha}f_{,\mu}). \label{eq:tor trans}%
\end{equation}
Conversely, under both the conformal transformation of the metric
(\ref{eq:conf trans}) and the transformation of torsion (\ref{eq:tor trans})
the connection transforms according to an Einstein's $\lambda$-transformations
(\ref{eq:conn trans}).Henceforth we will refer to (\ref{eq:conf trans}) of
$g_{\mu\nu}$ and (\ref{eq:tor trans}) as \textit{Cartan transformations}.

Clearly, the components of curvature tensor of the Riemann-Cartan space-time
are left invariant under Cartan transformations, that is, $\overline
{R}_{\;\:\beta\mu\nu}^{\,\alpha}=R_{\;\:\beta\mu\nu}^{\alpha}.$ On the other
hand, it is evident that the Ricci tensor is also invariant, i.e.
$\overline{R}_{\beta\nu}=\overline{R}_{\;\:\beta\lambda\nu}^{\,\lambda
}=R_{\beta\nu}$, while the curvature scalar transforms as $\overline{R}=
\overline{g}^{\mu\nu}\overline{R}_{\mu\nu}=e^{-f}R$. Finally, it follows from
the previous transformations that the Weyl tensor $W_{\:\:\beta\mu\nu}%
^{\alpha}=R_{\:\:\beta\mu\nu}^{\alpha}- \frac{1}{2}(\delta_{\:\:\mu}^{\alpha
}R_{\beta\nu} -\delta_{\:\:\nu}^{\alpha}R_{\beta\mu}-g_{\beta\mu}R_{\:\:\nu
}^{\alpha}+ g_{\beta\nu}R_{\:\:\mu}^{\alpha})-\frac{1}{6}(\delta_{\:\:\mu
}^{\alpha}g_{\beta\nu} -\delta_{\:\:\nu}^{\alpha}g_{\beta\mu})R$ is also left invariant.

As is known, the torsion tensor can be decomposed into three independent
irreducible parts: the tensorial, the vectorial and the axial-vectorial parts
\cite{Hehl,Capozziello2}. Here, we will consider the decomposition into only
two parts, namely, the vectorial torsion, which is determined by the torsion
trace defined by $\mathcal{\mathcal{T}_{\nu}=T}_{\;\:\:\alpha\nu}^{\alpha}$,
and the traceless part $L{}_{\;\:\mu\nu}^{\alpha}$ (including both the
tensorial and axi-vetorial parts). In this way, we have
\begin{equation}
\mathcal{T}{}_{\;\:\mu\nu}^{\alpha}=L_{\;\:\mu\nu}^{\alpha}+ \frac{1}%
{3}(\delta_{\;\:\mu}^{\alpha}\mathcal{T}_{\nu}- \delta_{\;\:\nu}^{\alpha
}\mathcal{T}_{\mu}). \label{eq:tor dec}%
\end{equation}
Using the above decomposition we can verify that, under a transformation given
by (\ref{eq:conf trans}) and (\ref{eq:conn trans}), the trace of the torsion
effectively plays the role of a gauge vector field, whereas the traceless part
is left invariant, that is,%

\begin{equation}%
\begin{array}
[c]{ccc}%
\overline{L}_{\:\:\mu\nu}^{\alpha}=L_{\:\:\mu\nu}^{\alpha}, & \;\overline
{\mathcal{T}}_{\mu} & =\mathcal{T}_{\mu}-\frac{3}{2}\, f_{,\mu}.
\end{array}
\label{eq:tor parts trans}%
\end{equation}

Let us now assume that the Riemann-Cartan space-time we are considering is
characterized by the following conditions: First, we require that the
traceless part of the torsion vanishes identically $L_{\;\:\mu\nu}^{\alpha}%
=0$. Second, let us suppose the existence of a scalar field $\phi(x)$, which
completely determines $\mathcal{T}_{\mu}$ through the equation
\begin{equation}
\mathcal{T}_{\mu}=-\frac{3}{2}\phi_{,\mu}, \label{eq:torsca trace}%
\end{equation}
that is, the torsion trace $\mathcal{T}_{\mu}$ is the gradient of a scalar
``potential'' $\phi(x)$ . This torsion determined by the gradient of a
propagating potential, also called gradient torsion\textit{ }%
\cite{gradtorsion}, will be referred to as \emph{scalar torsion}, while the
scalar field $\phi(x)$ will be called\textit{ torsion scalar field}. Thus, the
Riemann-Cartan space-time with scalar torsion is characterized by two
fundamental geometric objects, namely, the metric $g_{\mu\nu}$ and the torsion
scalar field $\phi$, which transforms under \ (\ref{eq:conf trans}) and
(\ref{eq:tor trans}) as
\begin{equation}%
\begin{array}
[c]{c}%
\begin{array}
[c]{ccc}%
\overline{g}_{\mu\nu} & = & e^{f}g_{\mu\nu},\;\overline{\phi}= \phi+f.
\end{array}
\end{array}
\label{eq:cart trans}%
\end{equation}
Since the transformations (\ref{eq:cart trans}) are mappings between
Riemann-Cartan space-times with scalar torsion, they will called
\textit{Cartan (gauge) transformations}. The set $(M,g,\phi)$ consisting of a
differentiable space-time manifold $M$ endowed with a Lorentzian metric $g$
and a torsion scalar field $\phi$ will be referred to as a \textit{Cartan
gauge}. Any Cartan gauge $(M,g,\phi)$ is related to another $(\overline
{M},\overline{g},\overline{\phi})$ by a Cartan transformation. It is easy to
see that by means of a Cartan transformation (\ref{eq:cart trans}) it is
possible to go from an arbitrary Cartan gauge $(M,g,\phi)$ to a unique Cartan
gauge $(M,\widehat{g},\widehat{\phi})=(M,\widehat{g},0)$, where $\widehat
{\phi}=0$, the latter being called \textit{Riemann gauge}, since in this gauge
the torsion vanishes. \footnote{From now on all quantities referred to the
Riemann gauge will denoted by a big hat superscript.}

As we will see in the next section, the Riemann gauge plays a fundamental role
in the formulation of general relativity in an arbitrary Riemann-Cartan
space-time with a scalar torsion. In fact, we will take advantage of a very
important property of the Riemann gauge, which will be used as a heuristic
tool to formulate general relativity in an arbitrary Riemann-Cartan space-time
with a scalar torsion. Consider a Cartan gauge $(M,g,\phi)$ and the unique
Riemann gauge $(M,\widehat{g},0)$, which is related to it by a Cartan
transformation. It is not difficult to verify that the metric $\widehat
{g}_{\mu\nu}$ and the connection $\widehat{\Gamma}_{\:\:\mu\nu}^{\,\alpha}=
\{_{\mu\nu}^{\,\alpha}\}_{\widehat{g}}= \frac{1}{2}\widehat{g}^{\,\alpha\beta
}(\widehat{g}_{\beta\nu,\mu}+ \widehat{g}_{\mu\beta,\nu}-\widehat{g}_{\mu
\nu,\beta})$ of the Riemann gauge $(M,\widehat{g},0)$ can be expressed in
terms of the metric $g_{\mu\nu}$ and the connection $\Gamma_{\;\mu\nu}%
^{\alpha}$ of the Cartan gauge $(M,g,\phi)$, respectively, through the relations%

\begin{equation}
\widehat{g}_{\mu\nu}=e^{-\phi}g_{\mu\nu}\:(\:\widehat{g}^{\,\alpha\beta}=
e^{\phi}g^{\alpha\beta})\,,\;\widehat{\Gamma}_{\:\:\mu\nu}^{\alpha}=
\Gamma_{\;\:\mu\nu}^{\alpha}-\frac{1}{2}\phi_{,\mu}\delta_{\;\:\nu}^{\alpha
}\,, \label{eq:invariant g gam}%
\end{equation}
which are invariant with respect to Cartan transformations. Therefore, these
relations allow one to express quantities calculated in the Riemann gauge in
terms of quantities in the Cartan gauge, without any reference to the Riemann
gauge. This is an important property of the Riemann gauge, since not only
$\widehat{g}_{\mu\nu}$ and $\widehat{\Gamma}_{\:\:\mu\nu}^{\alpha}$, but also
any quantity calculated by using them, will be invariant under Cartan
transformations. In fact, the relations (\ref{eq:invariant g gam}) provide a
prescription to define quantities in a Cartan gauge, starting from their
definitions in the related Riemann gauge, in a way that is clearly invariant
under Cartan transformations. They are also useful in clarifying the meaning
of some physical quantities and equations, whose behavior in a Cartan gauge
may appear rather different from their behavior in the Riemann gauge. As we
will see later, such is the case, for instance, of the conservation law of
matter energy-momentum tensor.

Following the line of thought outlined above, let us consider an arbitrary
Cartan gauge $(M,g,\phi)$, in which the scalar torsion and the corresponding
contortion will be given, respectively, by
\begin{equation}
\mathcal{T}{}_{\;\:\mu\nu}^{\alpha}=\frac{1}{2}(\delta_{\;\:\nu}^{\alpha}%
\phi{}_{,\mu} -\delta_{\;\:\mu}^{\alpha}\phi_{,\nu}), \label{eq:tor sca}%
\end{equation}
\begin{equation}
K{}_{\:\:\mu\nu}^{\alpha}=-\frac{1}{2}g^{\alpha\beta}(g_{\beta\mu}\phi{}%
_{,\nu} -g_{\mu\nu}\phi{}_{,\beta})=-\frac{1}{2}(\delta_{\:\:\mu}^{\alpha}%
\phi{}_{,\nu} -g_{\mu\nu}\phi^{,\alpha}). \label{eq:contor sca}%
\end{equation}
Under these assumptions, it is easy to verify that, in the gauge $(M,g,\phi)$,
the Ricci tensor may be written as
\begin{equation}
R_{\mu\nu}=\widetilde{R}_{\mu\nu}-(\widetilde{\nabla}_{\mu}\phi_{,\nu}+
\frac{1}{2}g_{\mu\nu}\widetilde{\Box}\phi)-\frac{1}{2}(\phi_{,\mu}\phi_{,\nu}-
g_{\mu\nu}\phi_{,\alpha}\phi^{,\alpha}), \label{eq:ricci ten}%
\end{equation}
and the curvature scalar will be given by
\begin{equation}
R=\widetilde{R}-3\widetilde{\Box}\phi+\frac{3}{2}\phi_{,\alpha}\phi^{,\alpha},
\label{eq:curv sca}%
\end{equation}
where $\widetilde{\Box}=g^{\alpha\beta}\widetilde{\nabla}_{\alpha}%
\widetilde{\nabla}_{\beta}$ . (Note that the Ricci tensor is symmetric, since
$\widetilde{\nabla}_{\mu}\phi_{,\nu}=\widetilde{\nabla}_{\nu}\phi_{,\mu}$). At
this point, let us recall that in a Riemann-Cartan manifold the Bianchi
identities are given by $R_{\:\:[\beta\mu\nu]}^{\alpha}+ \mathcal{T}%
_{\:\:\:\:[\beta\mu;\nu]}^{\alpha}+ \mathcal{T}_{\:\:\:\:[\beta\mu}^{\lambda
}\mathcal{T}_{\:\:\:\:\nu]\lambda}^{\alpha}=0$ and $R_{\:\:\beta[\mu\nu
;\sigma]}^{\alpha}+ R_{\:\:\beta\lambda[\mu}^{\alpha}\mathcal{T}%
_{\:\:\:\:\nu\sigma]}^{\lambda}=0$. In the case of scalar torsion, the above
identities reduce to $R_{\:\:[\beta\mu\nu]}^{\alpha}=0$ and $R_{\:\:\beta
[\mu\nu;\lambda]}^{\alpha}+R_{\:\:\beta[\mu\nu}^{\alpha}\phi_{,\lambda]}=0$.
Finally, the contracted Bianchi identities are given by $\nabla^{\alpha
}G_{\alpha\beta}+\phi^{,\alpha}G_{\alpha\beta}=0$, where $G_{\alpha\beta
}=R_{\alpha\beta}-\frac{1}{2}g_{\alpha\beta}R$ denotes the Einstein tensor of
the Riemann-Cartan space-time manifold. It is clear that the contracted
Bianchi identities can be written, alternatively, as
\begin{equation}
\nabla^{\alpha}(e^{\phi}G_{\alpha\beta})=0. \label{eq:bian div}%
\end{equation}
Of course all these identities are invariant under Cartan transformations
(\ref{eq:cart trans}). A simple way to show the above results is to start with
the Bianchi identities in the Riemann gauge $(M,\widehat{g},0)$, where the
torsion scalar field is null, and then express them in the Cartan gauge
$(M,g,\phi)$ by using the invariant relations (\ref{eq:invariant g gam})
between their metrics and connections, namely, $\widehat{g}_{\mu\nu}=e^{-\phi
}g_{\mu\nu}$ and $\widehat{\Gamma}_{\:\:\mu\nu}^{\alpha}=\Gamma_{\;\:\mu\nu
}^{\alpha}- \frac{1}{2}\phi_{,\mu}\delta_{\;\:\nu}^{\alpha}$.

We assume that physical quantities are invariant under Cartan transformations.
Thus, all Cartan gauges related by Cartan transformations may be viewed as
different representations of the same fundamental physical entity: the
gravitational field, bearing in mind that in an arbitrary Cartan gauge the
gravitational field is determined by two independent geometric objects,
namely, the metric and the torsion scalar field.

As we know, there are two types of distinguished \textit{unparametrized}
curves, whose properties are defined by the metric and a general linear
connection. The first type is the \textit{geodesic}, also called metric
geodesic, which has the property that the space-time interval $ds^{2}%
=g_{\mu\nu}dx^{\mu}dx^{\nu}$ is an extremum along the curve. Thus, a
parametrized time-like curve $x^{\mu}=x^{\mu}(\lambda)$, passing through the
events $x^{\mu}(a)$ and $x^{\mu}(b)$, corresponds to a geodesic if and only if
it extremizes the time-like space-time interval functional (considering metric
assignature -2)%

\begin{equation}
\Delta s=\int_{a}^{b}\left(  g_{\mu\nu}\frac{dx^{\mu}}{d\lambda} \frac
{dx^{\nu}}{d\lambda}\right)  ^{\frac{1}{2}}d\lambda. \label{eq: ds integ}%
\end{equation}
It should be noted that in spite of the fact that the time-like space-time
interval above reduces to the known expression of the proper time in general
relativity in the Riemann gauge, there is no reparametrization that could make
it invariant under Cartan transformations (\ref{eq:cart trans}). Therefore, we
cannot take $\Delta s$ as the extension to an arbitrary Cartan gauge of the
general relativity clock hypothesis, i.e., the assumption that $\Delta s$
measures the proper time interval measured by a clock attached to a test
particle. In the next section, we will show how to sort out this problem. It
is not difficult to verify that the extremization condition of the space-time
interval functional leads to the geodesic equations
\begin{equation}
\frac{d^{2}x^{\alpha}}{d\lambda^{2}}+\left\{  _{\mu\nu}^{\,\alpha}\right\}
\frac{dx^{\mu}}{d\lambda}\frac{dx^{\nu}}{d\lambda}=\frac{d^{2}x^{\alpha}%
}{d\lambda^{2}} +(\Gamma{}_{\:\:\mu\nu}^{\alpha}-K{}_{\:\:\mu\nu}^{\alpha})
\frac{dx^{\mu}}{d\lambda}\frac{dx^{\nu}}{d\lambda}=0, \label{eq: geod afpar}%
\end{equation}
where the parameter $\lambda$ is called an affine parameter, since it is fixed
up to an affine transformation $\lambda^{\prime}=a\lambda+b$ ($a,b$
constants). As in general relativity, the parameter $\lambda$ can be
identified with the space-time interval $s$ along the curve (up to affine
transformations) and, therefore, the quantity $(\frac{ds}{d\lambda}%
)^{2}=g_{\mu\nu}\frac{dx^{\mu}}{d\lambda} \frac{dx^{\nu}}{d\lambda}=(\frac
{1}{a})^{2}$ is constant under Riemann-Cartan parallel transport along the
geodesic curve, that is, $\frac{d}{d\lambda}(g_{\mu\nu}\frac{dx^{\mu}%
}{d\lambda} \frac{dx^{\nu}}{d\lambda})=\frac{dx^{\alpha}}{d\lambda}%
\nabla_{\alpha}(g_{\mu\nu} \frac{dx^{\mu}}{d\lambda}\frac{dx^{\nu}}{d\lambda
})=0$. Therefore, the equations of a geodesic curve $x^{\mu}=x^{\mu}(s)$, with
the space-time interval $s$ being used as a parameter and the tangent vector
$v^{\mu}=\frac{dx^{\mu}}{ds}$, are given by%

\begin{equation}
\frac{d^{2}x^{\alpha}}{ds^{2}}+\{{}_{\mu\nu}^{\,\alpha}\} \frac{dx^{\mu}}%
{ds}\frac{dx^{\nu}}{ds}=\frac{d^{2}x^{\alpha}}{ds^{2}}+ (\Gamma{}_{\:\:\mu\nu
}^{\alpha}-K{}_{\:\:\mu\nu}^{\alpha})\frac{dx^{\mu}}{ds}\frac{dx^{\nu}}{ds}=0.
\label{eq:geod ds}%
\end{equation}

Since the geodesics depend only on the metric, their Cartan transformations
(\ref{eq:cart trans}) involve only the conformal transformations of the
metric. Therefore, as in general relativity, they are not invariant under
these transformations. The only exception occurs with null geodesics, which
require a reparametrization \cite{Wald}.

The second type of unparametrized curves are the so-called
\textit{auto-parallel} curves, also called \textit{affine geodesics}, since
they are curves which parallel-transports their tangent vectors along
themselves with respect to the Riemann-Cartan connection \cite{Hehl,Schouten}.
In general the equations of an auto-parallel curve $x^{\mu}=x^{\mu}(\sigma)$ ,
with parameter $\sigma$ and tangent vector $\frac{dx^{\mu}}{d\sigma}$ , are
given by
\begin{equation}
\frac{D}{D\sigma}(\frac{dx^{\alpha}}{d\sigma})= \frac{dx^{\mu}}{d\sigma}%
\nabla_{\mu}(\frac{dx^{\alpha}}{d\sigma})= \frac{d^{2}x^{\alpha}}{d\sigma^{2}%
}+\Gamma{}_{\:\:\mu\nu}^{\alpha} \frac{dx^{\mu}}{d\sigma}\frac{dx^{\nu}%
}{d\sigma}=h(\sigma) \frac{dx^{\alpha}}{d\sigma}, \label{eq:autopar gen}%
\end{equation}
where $h(\sigma)$ is a function of the curve parameter $\sigma$. The function
$h(\sigma)$ can be removed by a reparametrization $\sigma=\sigma(\lambda)$, so
that the equations of an auto-parallel curve can be written as
\begin{equation}
\frac{d^{2}x^{\alpha}}{d\lambda{}^{2}}+\Gamma{}_{\:\:\mu\nu}^{\alpha}
\frac{dx^{\mu}}{d\lambda}\frac{dx^{\nu}}{d\lambda}=0, \label{eq:autopar afpar}%
\end{equation}
and the new parameter $\lambda$ is an affine parameter. The affine parameter
$\lambda$ of an auto-parallel curve can also be identified with the space-time
interval $s$, up to an affine transformation, since the quantity $(\frac
{ds}{d\lambda})^{2}= g_{\mu\nu}\frac{dx^{\mu}}{d\lambda}\frac{dx^{\nu}%
}{d\lambda}$ is also constant under Riemann-Cartan parallel transport along
the auto-parallel curve. Therefore, the equations of an auto-parallel curve
parametrized by the space-time interval, with $v^{\alpha}=\frac{dx^{\alpha}%
}{ds}$, can be written as
\begin{equation}
\frac{Dv^{\alpha}}{Ds}=v^{\mu}\nabla_{\mu}v^{\alpha}= \frac{dv^{\alpha}}%
{ds}+\Gamma{}_{\:\:\mu\nu}^{\alpha}v^{\mu}v^{\nu}=0. \label{eq:autopar ds}%
\end{equation}

The auto-parallel curves, parametrized either by the space-time interval
(\ref{eq:autopar ds}) or by an arbitrary affine parameter
(\ref{eq:autopar afpar}), are invariant under Cartan transformations
(\ref{eq:cart trans}), although they need to be reparametrized. Due to the
identification of the affine parameter with the space-time interval, the
necessary reparametrization of the affine parameter must be equal to the
space-time interval transformation $d\overline{s}=e^{\frac{1}{2}f}ds$, that
is, $d\overline{\lambda}=e^{\frac{1}{2}f}d\lambda$. Therefore, they are the
natural candidates to describe the motion of test particles and light rays
which, are assumed to be invariant under Cartan transformations, as we will
see in the next section. In general, geodesics and auto-parallel curves are
different unparametrized curves in a Riemann-Cartan space-time manifold, in
the sense that geodesics equations cannot be transformed into auto-parallels
equations through a reparametrization. There are only two exceptions. The
first exception occurs with null geodesics and null auto-parallels, whose
equations can be transformed one into another by means of a reparametrization.
The second appears when the torsion is totally antisymmetric (axial torsion),
since the contortion is also totally antisymmetric and, therefore, geodesic
and auto-parallels coincide \cite{Rodrigues}.

Finally, let us remark that geodesics plays a fundamental role in general
relativity as well as in most metric theories of gravity. Indeed, an elegant
aspect of the geometrization of the gravitational field lies in the geodesics
postulate, i.e., the statement that light rays and test particles moving under
the influence of gravity alone follow space-time geodesics. Therefore a great
deal of information about the motion of particles in a given space-time is
promptly available once one knows its geodesics.

The fact that auto-parallel curves are invariant under Cartan gauge
transformations and that Riemannian geometry can be seen as a particular case
of Riemann-Cartan geometry seems to suggest that it should be possible to
express general relativity in a more general geometrical setting, namely, one
in which not only the motion of particles and light rays but also the form of
the gravitational field equations are also invariant under Cartan gauge transformations.

\section{General Relativity and a New Kind of Invariance}

In this section we will show that it is possible to express general relativity
in a Riemann-Cartan space-time framework. This program will involve several
independent steps, consisting basically of the following points: the
formulation of the gravitational field equations through a variational
principle, the determination of the motion of test particles and light rays,
the coupling between matter fields and the gravitational field in a Cartan
gauge and, finally, the energy-momentum conservation. We will also include a
further requirement: invariance under Cartan transformations. As we will see,
the formulation of general relativity in a special class of Riemann-Cartan
space-times (those endowed with scalar torsion) entails a complete equivalence
among all Cartan gauges. This a consequence of the fact that the mentioned
formulation is invariant under Cartan transformations, with the usual
Riemannian formulation of general relativity being restored in the Riemann
gauge, where the torsion scalar field vanishes. In a certain sense, this
reminds us of the gauge freedom exhibited by classical electrodynamics.

To carry out the program outlined above, let us start with the Lagrangian
formulation of general relativity. Consider the Einstein-Hilbert Lagrangian
$L_{EH}=\sqrt{-\widehat{g}}\,\widehat{R}$ in a Riemann gauge. In order to
obtain the same Lagrangian expressed in an arbitrary Cartan gauge we will make
use of (\ref{eq:invariant g gam}) . This will lead to $L_{EH}=\sqrt
{-g}e^{-2\phi}(e^{\phi}R)=\sqrt{-g}e^{-\phi}R$. If we want to include the
cosmological constant and matter, then the simplest action that can be built
under the above conditions is
\begin{equation}
S=\int d^{4}x\sqrt{-g}e^{-2\phi}(e^{\phi}R+2\Lambda+\kappa L_{m}),
\label{eq: S action}%
\end{equation}
where $L_{m}$ stands for the Lagrangian of the matter fields, $\Lambda$ is the
cosmological constant and $\kappa$ is the Einstein`s constant. Let us remark
that the invariance of the action with respect to Cartan transformations
requires the same invariance of both $\Lambda$ and $L_{m}$. The gravitational
field equations in a Cartan gauge are obtained by requiring the action
(\ref{eq: S action}) to be stationary under arbitrary variations with respect
to the two independent elements of the gravitational field, namely, the metric
and the torsion scalar field. In this way, we will get the following
equations:
\begin{equation}
G_{\mu\nu}[g,\phi]=\widetilde{G}_{\mu\nu}-(\widetilde{\nabla}_{\mu}\phi_{,\nu
}- g_{\mu\nu}\widetilde{\Box}\phi)-\frac{1}{2}(\phi_{,\mu}\phi_{,\nu}+
\frac{1}{2}g_{\mu\nu}\phi_{,\alpha}\phi^{,\alpha})=-\kappa T_{\mu\nu}+
g_{\mu\nu}e^{-\phi}\Lambda, \label{eq:fieldeq g}%
\end{equation}
and
\begin{equation}
R[g,\phi]=\widetilde{R}-3\widetilde{\Box}\phi+\frac{3}{2}\phi_{,\alpha}%
\phi^{,\alpha}= \kappa T-4e^{-\phi}\Lambda, \label{eq: fieldeq phi}%
\end{equation}
where $T=g^{\mu\nu}T_{\mu\nu}$ is the trace of the energy-momentum tensor of
matter $T_{\mu\nu}$ \footnote{Here we are denoting the Einstein tensor and the
curvature scalar of the Riemann-Cartan space-time by $G_{\mu\nu}[g,\phi]$ and
$R[g,\phi]$, respectively, to emphasize that they depend both on the metric
$g$ and on the torsion scalar field $\phi$. Whenever necessary, the same
notation will be used for other quantities of the Riemann-Cartan space-time.}.
It should be noted that the above gravitational field equations are not
independent, since (\ref{eq: fieldeq phi}) is just the trace of
(\ref{eq:fieldeq g}). This is consistent with the fact that we have complete
freedom in the choice of the Cartan gauge through the Cartan transformations
(\ref{eq:cart trans}). It also means that one degree of freedom in the pair
$(g_{\mu\nu},\phi)$ may be viewed as an arbitrary gauge field and not as a
dynamical field. Each Cartan gauge corresponds to a different choice of gauge,
where either $\phi$ or one of the independent component of $g_{\mu\nu}$ is
chosen arbitrarily. Therefore, the gravitational field in a Cartan gauge has
the same number of degrees of freedom as the gravitational field in usual
formulation of general relativity.

Our next task it to extend Einstein's geodesic postulate to arbitrary Cartan
gauges, in an invariant way under Cartan transformations. It will be required
that in the Riemann gauge the motion of test particles and light rays in the
Riemann gauge should be governed by Einstein's geodesic postulate. It is
almost obvious that to complete the required extension we need the following
statement: If we represent parametrically a time-like curve as $x^{\mu}%
=x^{\mu}(\lambda)$, then this curve will represent the world line of a test
particle free from all non-gravitational forces, passing through the events
$x^{\mu}(a)$ and $x^{\mu}(b)$, if and only if it extremizes the functional
\begin{equation}
\Delta\tau=\int_{a}^{b}(e^{-\phi}g_{\mu\nu}\frac{dx^{\mu}}{d\lambda}
\frac{dx^{\nu}}{d\lambda})^{\frac{1}{2}}d\lambda. \label{eq: prop time}%
\end{equation}
Clearly, the above definition is obtained from the special relativistic
expression of proper time by using the coupling prescription $\eta_{\mu\nu
}\rightarrow e^{-\phi}g_{\mu\nu}.$ The right-hand side of this equation is
invariant under Cartan transformations and reduces to the known expression of
the proper time in general relativity in the Riemann gauge. Therefore, we take
$\Delta\tau$, as given above, as the extension to an arbitrary Cartan gauge of
the clock hypothesis, i.e., the assumption that $\Delta\tau$ measures the
proper time measured by a clock attached to the test particle. Note that, in a
Cartan gauge, the proper time depends on both fundamental elements of the
gravitational field, i.e., $(g_{\mu\nu},\phi)$. It is not difficult to verify
that the extremization condition of the proper time functional
(\ref{eq: prop time}) leads to the following equations of motion of a test
particle moving only under the gravitational field influence
\begin{equation}
\frac{d^{2}x^{\alpha}}{d\lambda^{2}}+\{_{\mu\nu}^{\,\alpha}\} \frac{dx^{\mu}%
}{d\lambda}\frac{dx^{\nu}}{d\lambda}+ \frac{1}{2}\phi^{,\beta}g{}_{\mu\nu
}\frac{dx^{\mu}}{d\lambda} \frac{dx^{\nu}}{d\lambda}-\phi_{,\nu}\frac{dx^{\nu
}}{d\lambda} \frac{dx^{\alpha}}{d\lambda}=0. \label{eq:motion afpar}%
\end{equation}
Let us recall that in the derivation of the above equations it was considered
the fact that the quantity $(\frac{d\tau}{d\lambda})^{2}= e^{-\phi}g_{\mu\nu
}\frac{dx^{\mu}}{d\lambda}\frac{dx^{\nu}}{d\lambda}$ is constant under
Riemann-Cartan parallel transport along the curve, i.e., that $\frac{dx^{\mu}%
}{d\lambda}\nabla_{\mu}(e^{-\phi}g_{\mu\nu} \frac{dx^{\mu}}{d\lambda}%
\frac{dx^{\nu}}{d\lambda})=0$, which allows the identification of $\lambda$
with the proper time $\tau$, up to an affine transformation. Substituting the
expression for the contortion given by (\ref{eq:contor sca}) in
(\ref{eq:motion afpar}) we have
\begin{equation}
u^{\mu}\nabla_{\mu}u^{\alpha}-\frac{1}{2}\frac{d\phi}{d\tau}u^{\alpha}=
\frac{du{}^{\alpha}}{d\tau}+\Gamma{}_{\:\:\mu\nu}^{\alpha}u^{\mu}u^{\nu}
-\frac{1}{2}\frac{d\phi}{d\tau}u^{\alpha}=0, \label{eq:motion prtime}%
\end{equation}
where $u^{\beta}=\frac{dx^{\beta}}{d\tau}$ is the 4-velocity of the test
particle and $\frac{d\phi}{d\tau}=\phi_{,\mu}u^{\mu}$ is the variation of the
torsion scalar field along the curve. These are the equations of the
auto-parallel curves (\ref{eq:autopar gen}) defined by the Riemann-Cartan
connection, parametrized with the proper time (\ref{eq: prop time}). They are
invariant under Cartan transformations and reduce to the known geodesic
equations in general relativity in the Riemann gauge. On the other hand, the
proper time is not an affine parameter. Nevertheless, the standard form of the
equations of auto-parallel curves (\ref{eq:autopar ds}), with the space-time
interval as an affine parameter, can be obtained from (\ref{eq:motion prtime})
through a reparametrization $\tau=\tau(s)$ given by $d\tau=e^{-\frac{1}{2}%
\phi}ds$, which leads to $u^{\mu}\nabla_{\mu}u^{\alpha}-\frac{1}{2}\frac
{d\phi}{d\tau}u^{\alpha}= e^{\phi}v{}^{\mu}\nabla_{\mu}v^{\alpha}=0$, where
$v^{\alpha}=\frac{dx^{\alpha}}{ds}$. Therefore, the extension of the geodesic
postulate to a Cartan gauge by requiring the proper time functional
(\ref{eq: prop time}) to be an extremum along the path of a test particle free
from all non-gravitational forces, is equivalent to postulating that the
motion of a test particle on the gravitational field must follow auto-parallel
time-like curves defined by the Riemann-Cartan connection.

Considering the unusual form of the equations of motion
(\ref{eq:motion prtime}) of a test particle moving only under the
gravitational field influence, written in terms of the proper time $\tau$, it
is valuable to make some comparisons between the proper time $\tau$ and the
space-time interval $s$. The proper-time (\ref{eq: prop time}) is invariant
under Cartan transformations, but it is not an affine parameter for test
particle paths, that is, the auto-parallel time-like curves. On the other
hand, the space-time interval is an affine parameter for the test particles
paths, but it is not invariant under Cartan transformations. At this point it
is important to note that the tangent vector to the time-like auto-parallel
curves may have different behaviors, depending on whether the space-time
interval or the proper time is used as parameter. When the parameter is the
space-time interval, the norm of the tangent vector $v^{\alpha}=\frac
{dx^{\alpha}}{ds}$ is constant since $g_{\alpha\beta}v^{\alpha}v^{\beta}=1$
along the time-like auto-parallel curve. Indeed, we have $\frac{d}%
{ds}(g_{\alpha\beta}v{}^{\alpha}v{}^{\beta})=0$, which is a consequence of the
equations (\ref{eq:autopar ds}) of the auto-parallel curve with the space-time
interval as affine parameter. On the other hand, when the parameter is the
proper time, the norm of the 4-velocity vector $u^{\alpha}=\frac{dx^{\alpha}%
}{d\tau}$ depends on the value of the torsion scalar field, since $e^{-\phi
}g_{\alpha\beta}u{}^{\alpha}u{}^{\beta}=1$ is constant along the time-like
auto-parallel curve, according to $\frac{d}{d\tau}(e^{-\phi}g_{\alpha\beta}%
u{}^{\alpha}u{}^{\beta})= u^{\mu}\nabla_{\mu}(e^{-\phi}g_{\alpha\beta
}u^{\alpha}u^{\beta})=0$, as a consequence of the equations
(\ref{eq:motion prtime}) of the auto-parallel curve in which the affine
parameter is the proper time. We see, then, that the 4-velocity $u^{\alpha
}=\frac{dx^{\alpha}}{d\tau}$ differs from the tangent vector $v^{\alpha}%
=\frac{dx^{\alpha}}{ds}$ by a factor due to the torsion scalar field, that is,
$u^{\alpha}= e^{-\frac{1}{2}\phi}v^{\alpha}$, since the proper time in a
Cartan gauge depends on both the metric and the torsion scalar field. The
space-time interval, in turn, depends only on the metric. Owing to this
correction, the norm of the 4-velocity depends on the torsion scalar field
through the invariant expression $g_{\alpha\beta}u^{\alpha}u^{\beta}=e^{\phi}%
$. The norm of the 4-velocity will be constant only in the special case in
which the torsion scalar field is constant along the auto-parallel curve, that
is, when $\frac{d\phi}{d\tau}=\phi_{,\mu}u^{\mu}=0$. Indeed, in this case we
have $\frac{d}{d\tau}(g_{\alpha\beta}u^{\alpha}u{}^{\beta})= u^{\mu}%
\nabla_{\mu}(g_{\alpha\beta}u^{\alpha}u^{\beta})=e^{\phi}\frac{d\phi}{d\tau
}=0$.

A simple way to show the above results is to start from the Riemann gauge
$(M,\widehat{g},0)$, where $\widehat{\phi}=0,$ and then change to a Cartan
gauge $(M,g,\phi)$ by using the invariant relations (\ref{eq:invariant g gam}%
), that is, $\widehat{g}_{\mu\nu}=e^{-\phi}g_{\mu\nu}$ and $\widehat{\Gamma
}_{\:\:\mu\nu}^{\alpha}=\Gamma_{\;\:\mu\nu}^{\alpha} -\frac{1}{2}\phi_{,\mu
}\delta_{\;\:\nu}^{\alpha}$ . Thus, let us consider a test particle moving
only under the influence of the gravitational field. Considering that the
4-velocity $u^{\alpha}= \frac{dx^{\alpha}}{d\tau}$ is an invariant (the
coordinates and the proper time are both invariant) and that the norm of the
4-velocity $\widehat{u}^{\alpha}$ in the Riemann gauge is constant, it follows
that the norm of the 4-velocity in a Cartan gauge depends on the torsion
scalar field according to the invariant relation $\widehat{g}_{\alpha\beta
}\widehat{u}^{\alpha}\widehat{u}^{\beta}= e^{-\phi}g_{\alpha\beta}u^{\alpha
}u^{\beta}=1$. Moreover, since in the Riemann gauge the 4-acceleration
$\widehat{a}^{\beta}$ vanishes, in accordance with the principle of
equivalence, it follows that the test particle moves, in a Cartan gauge, along
a time-like auto-parallel curve with the proper time as an affine parameter.
Indeed, we have the invariant relations $\widehat{a}^{\beta}= \widehat{u}%
^{\mu}\widehat{\nabla}_{\mu}\widehat{u}^{\beta}= u^{\mu}\nabla_{\mu}u^{\beta
}-\frac{1}{2}\frac{d\phi}{d\tau}u^{\beta}= e^{\phi}v{}^{\mu}\nabla_{\mu
}v^{\alpha}=0$. These are the same equations of motion (\ref{eq:motion prtime}
) that were previously obtained from the extension of the geodesic postulate
and the definition of proper time in a Cartan gauge. Therefore, it seems
natural to define the 4-acceleration in a Cartan gauge as the invariant
quantity
\begin{equation}
a^{\beta}=u^{\mu}\nabla_{\mu}u^{\beta}-\frac{1}{2}\frac{d\phi}{d\tau}u^{\beta
}= e^{\phi}v{}^{\mu}\nabla_{\mu}v^{\alpha}.
\end{equation}
Two comments are in order. First, as a consequence of the dependence of the
norm of the 4-velocity on the torsion scalar field, given by $g_{\alpha\beta
}u^{\alpha}u^{\beta}=e^{\phi}$, it follows that the invariant relation
$g_{\alpha\beta}u^{\alpha}a^{\beta}=0$ is satisfied. Second, the condition
$a^{\beta}=0$, valid for a test particle moving only under the gravitational
field influence, is a consequence of the equivalence principle, and means that
the test particle follows a time-like auto-parallel curve parametrized with
proper time.

As we know, the geodesic postulate in general relativity is a statement which
rules not only to the motion of test particles, but also to the propagation of
light rays in space-time. Because the path of light rays are null curves, one
cannot use the proper time as a parameter to describe them. On the other hand,
it seems natural to assume that light rays follow null auto-parallel curves.
Thus, these curves cannot be defined in terms of the proper time functional
(\ref{eq: prop time}), which is identically null for these curves. They must
be characterized instead by their behavior with respect to the Riemann-Cartan
parallel transport. We will extend this postulate by simply assuming that
light rays follow null auto-parallel curves defined by the Riemann-Cartan
connection. The equations of motion of light rays are the equations of null
auto-parallel curves $x^{\mu}=x^{\mu}(\lambda)$ with affine parameter
$\lambda$. They can be obtained from the auto-parallel equations
(\ref{eq:autopar afpar}) by requiring the tangent vector to be a null vector,
that is, $g{}_{\mu\nu}\frac{dx^{\mu}}{d\lambda}\frac{dx^{\nu}}{d\lambda}=0$,
and are given by
\begin{equation}
\frac{d^{2}x^{\alpha}}{d\lambda{}^{2}}+\{_{\mu\nu}^{\,\alpha}\} \frac{dx^{\mu
}}{d\lambda}\frac{dx^{\nu}}{d\lambda}- \frac{1}{2}\phi_{,\nu}\frac{dx^{\nu}%
}{d\lambda}\frac{dx^{\alpha}}{d\lambda}=0 \label{eq: light ray afpar}%
\end{equation}
which, through a reparametrization $\lambda=$$\lambda(\sigma)$ given by
$d\lambda=e^{\frac{1}{2}\phi}d\sigma$, are equal to the equations of a null
(metric) geodesic (\ref{eq: geod afpar}) with affine parameter $\sigma$, that
is, $\frac{d^{2}x^{\alpha}}{d\sigma{}^{2}}+\{_{\mu\nu}^{\,\alpha}\}
\frac{dx^{\mu}}{d\sigma}\frac{dx^{\nu}}{d\sigma}=0$. Therefore, in a
Riemann-Cartan space-time with scalar torsion, null geodesics and null
auto-parallel curves coincide, up to a reparametrization, and are given by the
same unparametrized null curves. Consequently, in the Riemann-Cartan
space-time, these curves define the same light cone and the same local causal
structure. Furthermore, according to this result, the local causal structure
of space-time in an arbitrary Cartan gauge is determined only by the metric
structure (with an indirect influence of the torsion through the gravitational
field equations).

Note that the light cone structure of the space-time does not change under
Cartan transformations, since it is well known that the equations of null
geodesics are preserved under conformal transformations (\ref{eq:conf trans}),
although one needs to reparametrize the curves in the new gauge \cite{Wald}.
In the case of Riemann-Cartan space-times with scalar torsion, the invariance
of the equations of null auto-parallel curves also needs a reparametrization,
since the connection components $\Gamma_{\alpha\beta}^{\mu}$ change under
Cartan transformations (\ref{eq:cart trans}) according to (\ref{eq:conn trans}%
). As a consequence, the causal structure of space-time remains unchanged in
all Cartan gauges.

The following step in our formulation has a subtle point in the framework of
non-Riemannian space-times. It is related to the question of how to define the
coupling between the matter fields and the gravitational field in a Cartan
gauge. Clearly, the choice of this coupling must be guided by the requirement
of invariance of the field equations with respect to Cartan transformations.
It will be assumed that the matter Lagrangian $L_{m}$ in a Cartan gauge
generally depends on the two independent fundamental elements of the
gravitational field $g_{\mu\nu}$ and $\phi$ and on the matter fields, here
generically denoted by $\xi$. The Lagrangian $L_{m}[g,\phi,\xi,\nabla\xi]$ of
the matter field $\xi$ in a Cartan gauge can be obtained from the Lagrangian
of the matter field in special relativity theory $L_{\;\: m}^{sr}=L_{\;\:
m}^{sr}[\eta,\xi,\partial\xi]$, where $\eta=(\eta_{\mu\nu})$ is the Minkowski
metric, through the prescription%

\begin{equation}
\eta_{\mu\nu}\rightarrow\widehat{g}_{\mu\nu},\;\;\partial_{\mu}\rightarrow
\widehat{\nabla}_{\mu}, \label{eq:coupling geo}%
\end{equation}
where $\widehat{g}_{\mu\nu}=e^{-\phi}g_{\mu\nu}$$\:$($\widehat{g}^{\alpha
\beta}= e^{\phi}g^{\alpha\beta}$) and $\widehat{\nabla}_{\mu}$ is the
covariant derivative with respect to the connection $\widehat{\Gamma}%
_{\:\:\mu\nu}^{\alpha}=\Gamma_{\;\:\mu\nu}^{\alpha} -\frac{1}{2}\phi_{,\mu
}\delta_{\;\:\nu}^{\alpha}$. Therefore, given the Lagrangian of matter in
special relativity theory $L_{\;\: m}^{sr}=L_{\;\: m}^{sr}[\eta,\xi
,\partial\xi]$, we define the coupling between the matter fields and the two
independent elements of the gravitational field $(g_{\mu\nu},\phi)$ in a
Cartan gauge by the rule%

\begin{equation}
L_{\;\: m}^{sr}[\eta,\xi,\partial\xi]\rightarrow L_{m}[g,\phi,\xi,\nabla
\xi]=L_{\;\: m}^{sr}[\widehat{g},\phi,\xi,\widehat{\nabla}\xi]\,.
\label{eq:coupling mat}%
\end{equation}
Besides, in order to assure the invariance of the coupling procedure and the
matter Lagrangian $L_{m}[g,\phi,\xi,\nabla\xi]$ with respect to Cartan
transformations (\ref{eq:cart trans}), it is necessary to assume that the
matter field $\xi$ is also invariant, since by the above rule $L_{m}%
[g,\phi,\xi,\nabla\xi]$ depends on $g_{\mu\nu}$ and $\phi$ only through the
invariant combinations $e^{-\phi}g_{\mu\nu}$ and $e^{\phi}g^{\mu\nu}$.
Furthermore, it follows that the part of the action (\ref{eq: S action}) that
is responsible for the coupling of matter with the gravitational field is also
invariant. Finally, this prescription reduces to the principle of minimal
coupling adopted in general relativity when we set $\phi=0$, that is, in the
Riemann gauge.

The final step is closely related to the previous one, since it is the
definition of the energy-momentum tensor of the matter fields and the
formulation of its conservation law. Similarly to the Einstein equations in a
Riemann gauge, the gravitational field equations $G_{\mu\nu}=-kT_{\mu\nu}$
(\ref{eq:fieldeq g}) in a Cartan gauge has a geometric quantity $G_{\mu\nu}$
on the left-hand side and a non-geometric quantity $T_{\mu\nu}$ on the
right-hand side. On the other hand, the invariance of the field equations
(\ref{eq:fieldeq g}) under Cartan transformations require that both $G_{\mu
\nu}$ and $T_{\mu\nu}$ have the same transformation laws, in spite of being of
completely different nature. This condition put several restrictions to the
possible choices on the coupling between gravitation and matter and,
consequently, on the forms of both the Lagrangian and the energy-momentum
tensor of matter fields. A important restriction is the requirement of
invariance of the covariant components $T_{\mu\nu}$ with respect to Cartan
transformations, since $G_{\mu\nu}$ is already invariant under these
transformations. However, as we will show in the following, it is possible to
give a definition of the energy-momentum tensor, based on the previous
definition of the matter Lagrangian in a Cartan gauge, which satisfies these
conditions and is consistent with all previous steps.

The Lagrangian $L_{m}[g,\phi,\xi,\nabla\xi]$ of the matter field, defined in
an arbitrary Cartan gauge $(M,g,\phi)$ by the rule (\ref{eq:coupling mat}),
has a remarkable property: it depends on the basic elements of the
gravitational field in a Cartan gauge, $g_{\mu\nu}$ and $\phi$, only through
the invariant combinations $e^{-\phi}g_{\mu\nu}$ and $e^{\phi}g^{\mu\nu}$.
Therefore, it is natural to define the energy-momentum tensor $T_{\mu\nu
}[g,\phi,\xi,\nabla\xi]$ in a Cartan gauge by the formula
\begin{equation}
\delta S_{m}=\delta\int d^{4}x\sqrt{-g}e^{-2\phi}L_{m}[g,\phi,\xi,\nabla
\xi]=\int d^{4}x\sqrt{-g}e^{-2\phi}T_{\mu\nu}[g,\phi,\xi,\nabla\xi
]\delta(e^{\phi}g^{\mu\nu}), \label{eq: e-mom def}%
\end{equation}
where the variation must be carried out with respect to both $g_{\mu\nu}$ and
$\phi$. This definition is not only invariant under Cartan transformations
(\ref{eq:cart trans}), but also reduces to general relativity definition of
the energy-momentum tensor in the Riemann gauge. It should be mentioned that
the above definition is invariant under Cartan transformations, since besides
the invariants $L_{m}[g,\phi,\xi,\nabla\xi]$ and $e^{\phi}g^{\mu\nu}$ there is
also the invariant quantity $d\widehat{V}=\sqrt{-\widehat{g}}d^{4}x =\sqrt
{-g}e^{-2\phi}d^{4}x$.

We would like to conclude this section with a brief comment on the form of the
energy-momentum conservation law in a arbitrary Cartan gauge. We start with
the gravitational field equations in the Riemann gauge $(M,\widehat
{g},\widehat{\phi}=0)$, given by $\widehat{G}_{\mu\nu}[\widehat{g},0]=
-\kappa\widehat{T}_{\mu\nu}[\widehat{g},0,\xi,\widehat{\nabla}\xi]$, where the
Einstein tensor $\widehat{G}_{\mu\nu}[\widehat{g},0]$ is divergenceless. Due
to this property, it follows from (\ref{eq:fieldeq g}) that the
energy-momentum conservation law in the Riemann gauge $(M,\widehat{g}%
,\widehat{\phi}=0)$ is given by
\begin{equation}
\widehat{\nabla}^{\mu}\widehat{T}_{\mu\nu}= \widehat{g}^{\mu\alpha}%
\widehat{\nabla}_{\alpha}\widehat{T}_{\mu\nu}=0. \label{eq: e-mon div rie}%
\end{equation}
This shows that, in the Riemann gauge $(M,\widehat{g},\widehat{\phi}=0)$, the
matter field interacts only with the space-time metric, as it should be. On
the other hand, we obtain the same gravitational field equations
(\ref{eq:fieldeq g}), i.e., $G_{\mu\nu}[g,\phi]= -\kappa T_{\mu\nu}[g,\phi
,\xi,\nabla\xi]$, if we go to an arbitrary Cartan gauge $(M,g=e^{\phi}%
\widehat{g},\phi)$. However, now the Einstein tensor $G_{\mu\nu}[g,\phi]$ is
not divergenceless. Indeed, a straightforward calculation shows that
$\widehat{\nabla}^{\alpha}\widehat{G}_{\alpha\mu}=\nabla^{\alpha}(e^{\phi
}G_{\alpha\mu})=0$, in accordance with the contracted Bianchi identities
(\ref{eq:bian div}). Therefore, it follows that the energy-momentum
conservation law in the Cartan gauge $(M,g,\phi)$ takes the form
$\widehat{\nabla}^{\alpha}\widehat{T}_{\alpha\mu}=\nabla^{\alpha}(e^{\phi
}T_{\alpha\mu})=0$, whence%

\begin{equation}
\nabla^{\alpha}(e^{\phi}T_{\alpha\mu})=e^{\phi}(\nabla^{\alpha}T_{\alpha\mu}
+\phi^{,\alpha}T_{\alpha\mu})=e^{\phi}(\widetilde{\nabla}^{\alpha}T_{\alpha
\mu} -\phi^{,\alpha}T_{\alpha\mu}+\frac{1}{2}\phi_{,\mu}T)=0,
\label{eq: e-mon div car}%
\end{equation}
since $\nabla^{\alpha}T_{\alpha\mu}=\widetilde{\nabla}^{\alpha}T_{\alpha\mu}
-2\phi^{,\alpha}T_{\alpha\mu}+\frac{1}{2}\phi_{,\mu}T$. Note that the law of
conservation of the energy-momentum tensor (\ref{eq: e-mon div car}) is a
consequence of the Bianchi identities (\ref{eq:bian div}), not only in the
Riemann gauge but also in any Cartan gauge. Furthermore, it is not difficult
to verify that the above equations are invariant under the Cartan
transformations (\ref{eq:cart trans}). From the invariance of both the
gravitational field equations $G_{\mu\nu}=-\kappa T_{\mu\nu}$
(\ref{eq:fieldeq g}) and the Bianchi identities $\nabla^{\alpha}(e^{\phi
}G_{\alpha\mu})=0$, we obtain that in the Cartan gauge $(M,g,\phi)$ the law of
conservation of the energy-momentum tensor is given by $\nabla^{\alpha
}(e^{\phi}T_{\alpha\mu})=0$.

The presence of non-vanishing terms on the right-hand side of
(\ref{eq: e-mon div car}) may led someone to think that the energy-momentum is
not conserved in a Cartan gauge. However, we must remember that the torsion
scalar field $\phi$ is an intrinsic element of the gravitational field in the
Riemann-Cartan space-time and should necessarily appear in any equation
describing the behavior of matter in any Cartan gauge. This explain the
presence of $\phi$ coupled with $T_{\mu\nu}$ in (\ref{eq: e-mon div car}) and
the apparent violation of the energy-momentum conservation law. Note that the
familiar energy-momentum conservation equations of general relativity are
recovered when $\phi=0$, that is, in the Riemann gauge.

It is important to pay attention to the fact that all physical quantities are
invariant under Cartan transformations and only in the Riemann frame they are
given by the familiar general relativity expressions. In a Cartan frame, on
the other hand, they are given by more general expressions involving not only
the metric but also the torsion scalar and require a careful interpretation to
deal with the Riemann-Cartan framework. The unique Riemann gauge and all the
Cartan gauges are equivalent from the experimental point of view, since any
physical quantity has the same numerical value in any of them.

Let us now apply the above results to describe the interaction of gravity with
a perfect fluid in an arbitrary Cartan frame. The energy-momentum tensor of a
perfect fluid in special relativity is defined by $T^{\mu\nu}=(\rho
c^{2}+p)u^{\mu}u{}^{\nu}-p\eta^{\mu\nu}$, where $\rho$, $p$ and $u^{\mu}%
=\frac{dx^{\mu}}{d\tau}$ denotes, respectively, the proper density, the
pressure and the 4-velocity fields. We now need to express $T^{\mu\nu}$ in an
arbitrary Cartan gauge. Rewriting this expression as $T_{\mu\nu}=(\rho
c^{2}+p)\eta_{\mu\alpha}\eta_{\nu\gamma}u^{\alpha}u^{\gamma}-p\eta_{\mu\nu}$
and following the coupling prescription $\eta_{\mu\nu}\rightarrow e^{-\phi
}g_{\mu\nu}$, we obtain
\begin{equation}
T_{\mu\nu}=(\rho c^{2}+p)e^{-2\phi}g_{\mu\alpha}g_{\nu\gamma}u^{\alpha
}u^{\gamma} -pe^{-\phi}g_{\mu\nu}, \label{eq:e-mon fluid}%
\end{equation}
which is the desired expression of the energy-momentum tensor of a perfect
fluid in an arbitrary Cartan gauge. It is worth remembering that, due to the
field equations (\ref{eq:fieldeq g}), $T_{\mu\nu}$, as given by the above
equation, is required to be invariant under Cartan transformations, since the
Einstein tensor $G_{\mu\nu}$ is invariant. Therefore, it follows that $\rho$
and $p$ must be regarded as invariant quantities, since $u^{\alpha}%
=\frac{dx^{\alpha}}{d\tau}$ and $e^{-\phi}g_{\mu\nu}$ are invariant. The same
conclusion follows from the invariant relations $\widehat{T}_{\mu\nu}%
\widehat{u}^{\mu}\widehat{u}^{\nu}= T_{\mu\nu}u^{\mu}u^{\nu}=\rho c^{2}$ and
$\widehat{T}=\widehat{T}_{\mu\nu}\widehat{g}^{\mu\nu}= T_{\mu\nu}e^{\phi
}g^{\mu\nu}=e^{\phi}T=(\rho c^{2}-3p)$.

Let us now take a look at the motion of the fluid particles in an arbitrary
Cartan gauge. Considering, as source of the gravitational field, a
pressureless perfect fluid (dust), it is straightforward to show that the
fluid particles move along auto-parallel time-like curves as a consequence of
the energy-momentum conservation law (\ref{eq: e-mon div car}). This is in
accordance with the equivalence principle, now seen from the point of view of
a Cartan frame, through the extension of the geodesic postulate to a Cartan gauge.

Finally, let us consider the coupling between the gravitational and the
electromagnetic field in a Cartan gauge. In special relativity, the Lagrangian
of the electromagnetic field $F_{\mu\nu}=A_{\mu,\nu}-A_{\nu,\mu}$ is given by
$L_{\:\: em}^{sr}=\frac{1}{2}\eta^{\alpha\mu}\eta^{\beta\nu}F_{\alpha\beta
}F_{\mu\nu}$ \cite{Adler}. Applying the coupling prescription
(\ref{eq:coupling geo}), i.e., $\eta_{\mu\nu}\rightarrow\widehat{g}_{\mu\nu}$
and $\partial_{\mu}\rightarrow\widehat{\nabla}_{\mu}$, we obtain that the
Lagrangian in a Cartan gauge is given by $L_{em}=\frac{1}{2}e^{2\phi}%
g^{\alpha\mu}g^{\beta\nu}\widehat{F}_{\alpha\beta}\widehat{F}_{\mu\nu}$, where
the electromagnetic field is defined by $\widehat{F}_{\mu\nu}=\widehat{\nabla
}_{\nu}\widehat{A}_{\mu}-\widehat{\nabla}_{\mu}\widehat{A}_{\nu}$. As we have
seen, the assumption of invariance of the electromagnetic potential under
Cartan transformations $\widehat{A}_{\mu}=A_{\mu}$, also assures the
invariance of the Lagrangian of the electromagnetic field. At the first sight,
due to the above definition, it seems that the electromagnetic field couples
with the torsion scalar field in the Cartan gauge, but this is not true.
According to the coupling procedure (\ref{eq:coupling geo}), the covariant
derivative $\widehat{\nabla}_{\nu}\widehat{A}_{\mu}$ of the electromagnetic
potential used to define $\widehat{F}_{\mu\nu}$ is given by $\widehat{\nabla
}_{\nu}A_{\mu}=\widetilde{\nabla}_{\nu}A_{\mu} +\frac{1}{2}(\phi_{,\nu}A_{\mu
}+\phi_{,\mu}A_{\nu})- \frac{1}{2}g_{\mu\nu}\phi^{,\alpha}A_{\alpha}$, and,
thus, it follows that%

\begin{equation}
\widehat{F}_{\mu\nu}=\widehat{\nabla}_{\nu}\widehat{A}_{\mu} -\widehat{\nabla
}_{\mu}\widehat{A}_{\nu}=F_{\mu\nu}=A_{\mu,\nu} -A_{\nu,\mu}.
\label{eq:elet field car}%
\end{equation}
Because the electromagnetic field does not couple with the torsion scalar
field, the Lagrangian of the electromagnetic field in a Cartan gauge will be
given by%

\begin{equation}
L_{em}=\frac{1}{2}e^{2\phi}F^{\mu\nu}F_{\mu\nu}, \label{eq:elet lagr car}%
\end{equation}
where $F^{\mu\nu}=g^{\alpha\mu}g^{\beta\nu}F_{\alpha\beta}$. On the other
hand, since $\widehat{F}_{\mu\nu}=F_{\mu\nu}$ , it follows that $L_{em}$ is
also invariant under the gauge transformations of the electromagnetic
potential $A_{\mu}^{\prime}=A_{\mu}+f_{,\mu}$. Now, by applying the definition
of the energy-momentum tensor (\ref{eq: e-mom def}), we find that the
electromagnetic field energy-momentum tensor in a Cartan gauge is given by
\begin{equation}
T_{\mu\nu}=e^{\phi}(F_{\mu}^{\:\:\alpha}F_{\alpha\nu}+ \frac{1}{2}g_{\mu\nu
}F^{\alpha\beta}F_{\alpha\beta}), \label{eq:eq e-mon elect}%
\end{equation}
whose conservation law, according to (\ref{eq: e-mon div car}), is given by
\begin{equation}
\widetilde{\nabla}^{\mu}T_{\mu\nu}-\phi^{,\mu}T_{\mu\nu}+ \frac{1}{2}%
\phi_{,\nu}T=\widetilde{\nabla}^{\mu}(F_{\mu}^{\:\:\alpha}F_{\alpha\nu}
+\frac{1}{2}g_{\mu\nu}F^{\alpha\beta}F_{\alpha\beta})=0,
\label{eq:e-mon elec div}%
\end{equation}
and does not involve the torsion scalar field as well. Moreover, similarly to
general relativity, we obtain that light rays move on null metric geodesic in
a Cartan gauge, as a consequence of the conservation laws
(\ref{eq:e-mon elec div}). This is in accordance with the equivalence
principle and the extension of the geodesic postulate for the motion of light
rays in a Cartan gauge. This seems to complete our program of formulating
general relativity in a geometrical setting that exhibits a new kind of
invariance, namely, that with respect to Cartan transformations.

\section{The Newtonian limit in a general Cartan gauge}

In order to gain some insight into the meaning of this new representation of
general relativity developed in the previous sections, let us now proceed to
examine the Newtonian limit of general relativity in an arbitrary Cartan gauge
$(M,g,\phi)$. As we know, a metric theory of gravity is said to possess a
Newtonian limit in the non-relativistic weak-field regime if one can derive
Newton's second law from the geodesic equations and the Poisson equation from
the gravitational field equations. Let us see how general relativity, when
expressed in a form that is invariant under Cartan gauge transformations,
fulfills these requirements. The method we will employ here to treat this
problem is standard and can be found in most textbooks on general relativity
(see, for instance, \cite{Adler} ). Since in Newtonian mechanics the space
geometry is Euclidean, a weak gravitational field in a geometric theory of
gravity should manifest itself as a metric phenomenon\ through a slight
perturbation of the Minkowskian space-time metric. Thus, we consider a
time-independent metric tensor of the form
\begin{equation}
g_{\mu\nu}=\eta_{\mu\nu}+\epsilon h_{\mu\nu}, \label{quasi-Minkowskian}%
\end{equation}
where $n_{\mu\nu}$ is the Minkowski metric (with assignature -2), $\epsilon$
is a small parameter ($\epsilon^{2}\ll\epsilon$) and the term $\epsilon
h_{\mu\nu}$ represents a very small time-independent perturbation due to the
presence of some matter configuration. Because we are working in the
non-relativistic regime we will suppose that the velocity $V$ of the particle
along its path is much less than $c$, so that the parameter $\beta=\frac{V}%
{c}$ will be regarded as very small; hence in our calculations only
first-order terms in $\epsilon$ and $\beta$ will be retained. The same kind of
approximation will be assumed with respect to the torsion scalar field $\phi$,
which will be supposed to be static and small, i.e., of the same order as
$\epsilon$, and, in order to emphasize this fact, we will write $\phi
=\epsilon\varphi$, where $\varphi$ is a finite function. Then, adopting the
usual Minkowskian coordinates of special relativity we can write the line
element defined by (\ref{quasi-Minkowskian}) as $ds^{2}=(\eta_{\mu\nu
}-\epsilon h_{\mu\nu})dx^{\mu}dx^{\nu}$, which leads, in our approximation,
to
\begin{equation}
\frac{d\tau}{cdt}=e^{-\frac{1}{2}\phi}\frac{ds}{cdt}\cong(1 -\frac{1}%
{2}\epsilon\varphi)(1+\frac{1}{2}\epsilon h_{00})\cong[1 +\frac{1}{2}%
\epsilon(h_{00}-\varphi)],
\end{equation}
where $\tau$ is the proper time defined by (\ref{eq: prop time}), whose
inverse is given by
\begin{equation}
c\frac{dt}{d\tau}\cong[1-\frac{1}{2}\epsilon(h_{00}-\varphi)]. \label{dtdtau}%
\end{equation}
We will now consider, in the same approximation, the invariant affine geodesic
equations (\ref{eq:motion prtime}) , i.e.,
\begin{equation}
\frac{d^{2}x^{\mu}}{d\tau^{2}}+ \Gamma_{\;\:\alpha\beta}^{\mu}\frac
{dx^{\alpha}}{d\tau}\frac{dx^{\beta}}{d\tau}= \frac{1}{2}\frac{d\phi}{d\tau
}\frac{dx^{\mu}}{d\tau}, \label{affinegeodesics}%
\end{equation}
which, by changing the parameter from the proper time $\tau$ to time
coordinate $t$, can be written as
\[
\left(  \frac{dt}{d\tau}\right)  ^{2}\left(  \frac{d^{2}x^{\mu}}{dt^{2}}+
\Gamma_{\;\:\alpha\beta}^{\mu}\frac{dx^{\alpha}}{dt}\frac{dx^{\beta}}%
{dt}\right)  = \left[  -\frac{d}{dt}\left(  \frac{dt}{d\tau}\right)  +
\frac{1}{2}\frac{d\phi}{dt}\frac{dt}{d\tau}\right]  \left(  \frac{dt}{d\tau
}\right)  \frac{dx^{\mu}}{dt\:}.
\]
Since both $\frac{dt}{d\tau}$ , given by (\ref{dtdtau}), and $\phi
=\epsilon\varphi$ do not depend on the time coordinate, the right hand side of
the above equations is identically zero, and we obtain that
\begin{equation}
\frac{d^{2}x^{\alpha}}{dt^{2}}+\Gamma_{\;\:\mu\nu}^{\alpha}\frac{dx^{\mu}}%
{dt}\frac{dx^{\nu}}{dt}=0, \label{eq:affinegeodapprox}%
\end{equation}
recalling that the symbol $\Gamma_{\;\:\alpha\beta}^{\mu}$ designates the
components of the Riemann-Cartan affine connection with a scalar torsion. From
the Christoffel symbols and the contortion tensor (\ref{eq:contor sca}) it is
easy to verify that, to first order in $\epsilon,$ we have
\begin{equation}
\Gamma_{\;\:\mu\nu}^{\alpha}=\frac{\epsilon}{2}\eta^{\alpha\lambda}%
(h_{\lambda\nu,\mu}+ h_{\mu\lambda,\nu}-h_{\mu\nu,\lambda})- \frac{\epsilon
}{2}\eta^{\alpha\lambda}(\eta_{\lambda\mu}\varphi_{,\nu}- \eta_{\mu\nu}%
\varphi_{,\lambda}). \label{Cartanconnection2}%
\end{equation}
It is not difficult to see that, unless $\mu=\nu=0$, the product
$\Gamma_{\;\:\mu\nu}^{\alpha}\frac{dx^{\mu}}{dt}\frac{dx^{\nu}}{dt}$ is of
order $\epsilon\beta$ or higher. In this way, the affine geodesic equations
(\ref{eq:affinegeodapprox}) become, to first order in $\epsilon$ and $\beta$,
\begin{equation}
\frac{d^{2}x^{\alpha}}{dt^{2}}+c^{2}\Gamma_{\;\:00}^{\alpha}=0.
\label{equation-of-motion}%
\end{equation}
Clearly for $\alpha=0$ the equations (\ref{equation-of-motion}) reduces to an
identity. On the other hand, if $\alpha$ is a spatial index $i$, a simple
calculation yields $\Gamma_{\;\:00}^{i}=- \frac{\epsilon}{2}\eta^{ik}%
\frac{\partial}{\partial x^{k}}(h_{00}-\varphi)=\frac{\epsilon}{2}\delta
^{ik}\frac{\partial}{\partial x^{k}}(h_{00}-\varphi)$, hence the affine
geodesic equations (\ref{affinegeodesics}) in this approximation become, in
three-dimensional vector notation,
\begin{equation}
\frac{d\overrightarrow{V}}{dt}=-\frac{\epsilon c^{2}}{2}\overrightarrow
{\nabla}(h_{00}-\varphi),
\end{equation}
which is simply Newton's equations of motion in a classical gravitational
field, provided we identify the scalar gravitational potential with
\begin{equation}
U=\frac{\epsilon c^{2}}{2}(h_{00}-\varphi) \label{Newtonian-potential}%
\end{equation}
Note the presence of the torsion scalar field $\phi=\epsilon\varphi$ in the
above equation. It means that it is the combination $(\epsilon h_{00}-\phi)$
that represents the Newtonian potential in a Cartan gauge.

Let us now turn our attention to the Newtonian limit of the gravitational
field equations. For this purpose, it will be convenient to rewrite the field
equations (\ref{eq:fieldeq g}) with $\Lambda=0$ in the form
\begin{equation}
R_{\mu\nu}[g,\phi]=\widetilde{R}_{\mu\nu}-(\widetilde{\nabla}_{\mu}\phi_{,\nu
}+ \frac{1}{2}g_{\mu\nu}\widetilde{\Box}\phi)-\frac{1}{2}(\phi_{,\mu}%
\phi_{,\nu}- g_{\mu\nu}\phi_{,\alpha}\phi^{,\alpha})=-\kappa(T_{\mu\nu}%
-\frac{1}{2}g_{\mu\nu}T). \label{eq: fiedeq ric}%
\end{equation}
In the weak-field approximation, i.e., when $g_{\mu\nu}=\eta_{\mu\nu}+\epsilon
h_{\mu\nu}$ and $\phi=\epsilon\varphi$, it is easy to show that, to first
order in $\epsilon$, we have $\widetilde{R}_{00}=-\frac{1}{2}\nabla
^{2}\epsilon h_{00}$ and $\widetilde{\Box}\phi=-\nabla^{2}\epsilon\varphi$,
where $\nabla^{2}$ denotes the Laplacian operator calculated with the
Minkowski metric. On the other hand, because we are assuming a static regime
$\partial_{0}\phi=0$, the equation (\ref{eq: fiedeq ric}) for $\mu=\nu=0$ now
reads
\begin{equation}
-\nabla^{2}\left[  \frac{\epsilon}{2}(h_{00}-\varphi)\right]  =-\kappa(T_{00}
-\frac{1}{2}T). \label{Newton-Cartan}%
\end{equation}
Let us consider a configuration of matter distribution with low proper density
$\rho$ moving at non-relativistic speed. According to the previous section, in
an arbitrary Cartan gauge, the energy-momentum tensor of matter in this
configuration will be given by (\ref{eq:e-mon fluid}), that is, $T_{\mu\nu
}=(\rho c^{2}+p)e^{-2\phi}g_{\mu\alpha}g_{\nu\gamma}u^{\alpha}u^{\gamma}-
pe^{-\phi}g_{\mu\nu}$, where $\rho$ and $p$ are defined as invariant
quantities with respect to Cartan transformations. Putting $e^{-\phi}%
\simeq1-\epsilon\varphi$ and recalling that in a non-relativistic regime we
can neglect $p$ with respect to $\rho$, we will have $T_{00}=$ $T\simeq\rho
c^{2}$. In this way, we obtain, to first order in $\epsilon$, that $T_{\mu\nu
}\simeq\rho c^{2}\eta_{\mu\alpha}\eta_{\nu\gamma}u^{\alpha}u^{\gamma}$.
Finally, after substituting $\kappa=\frac{8\pi G}{c^{4}}$ into
(\ref{Newton-Cartan}), we obtain
\begin{equation}
\nabla^{2}\left[  \frac{\epsilon}{2}(h_{00}-\varphi)\right]  =\frac{4\pi
G}{c^{2}}\rho, \label{eq: poisson eq}%
\end{equation}
which corresponds to the Poisson equation for the gravitational field
$\nabla^{2}U=4\pi G\rho$, with the gravitational potential $U$ given by
(\ref{Newtonian-potential}).

\section{Different pictures of the same physical phenomena}

As we have shown in the previous section, when we go from one Cartan
gauge\ $(M,g,\phi)$ to another $(M,\overline{g},\overline{\phi})$ through a
Cartan transformation (\ref{eq:cart trans}), the gravitational field equations
and the paths of motion of test particles and light rays do not change. It is
important to note that the gravitational field in a Cartan gauge is described
by the Riemann-Cartan curvature tensor, which is invariant and depends on two
independent geometric elements: the metric and the torsion scalar field.
Furthermore, due to Cartan transformations, the metric properties are
decoupled from the parallel transport properties. Therefore, the same physical
phenomena are described by distinct geometrical pictures, which arise in
different Cartan gauges and require different physical interpretations related
to either the metric or the torsion or both. Nonetheless, the numerical values
of all physical quantities are the same in all gauges, since these quantities
are invariant under Cartan transformations. This property rises the question
about the existence of a gauge more suitable to perform the calculations of
some physical quantity.

In fact, each gravitational field is described by a member of an equivalence
class, the latter defined as the set of all Cartan gauges, with different
metrics and different connections. In this class, one metric can be obtained
from another and one connection can be obtained from another by means of a
Cartan transformation. On the other hand, each equivalence class is completely
determined by a solution of the Einstein equations in the Riemann gauge. The
fact that the same gravitational field is described by an equivalence class of
representations given by Cartan gauges may give rise to misinterpretations,
since we are used to regard the gravitational field as a manifestation of the
Riemannian curvature tensor, which depends only on the metric. This is
particularly evident in the two special cases considered below.

The first case leads to an important equivalence class of Riemann-Cartan
space-times. It is the class which characterizes the absence of gravitation,
given by Minkowsky space-time in the Riemann gauge $(M,\eta,0)$, where both
the curvature $\widehat{R}_{::\beta\mu\nu}^{\alpha}[\eta,0]=0$ and the torsion
$\widehat{T}_{::\mu\nu}^{\alpha}[\eta,0]=0$ vanish. Thus, in a Cartan gauge
$(M,e^{\phi}\eta,\phi)$, the metric is given by $g_{\mu\nu}=e^{\phi}\eta
_{\mu\nu}$ and the connection by $\Gamma_{\;:\mu\nu}^{\alpha}=\frac{1}{2}%
\phi_{,\mu}\delta_{\;:\nu}^{\alpha}$, where the torsion scalar field $\phi$ is
the conformal factor of the Minkowski metric, according to the invariant
relations (\ref{eq:invariant g gam}). These Riemann-Cartan space-times have
two features, which are invariant under Cartan transformations. They satisfy
the teleparallelism condition $R{}_{::\beta\mu\nu}^{\alpha}[e^{\phi}\eta
,\phi]=0$, and hence have a a Weitzenb\"{\i}\textquestiondown
${\frac12}$%
ck connection\cite{Aldrovandi}. They have a conformally flat metric, in the
sense that $g_{\mu\nu}=e^{\phi}\eta_{\mu\nu}$, defined by the invariant
condition $W{}_{::\beta\mu\nu}^{\alpha}[e^{\phi}\eta,\phi]=0$ (as in general
relativity). Therefore, the local causal structure is the same as the one of
Minkowski space-time. Furthermore, test particles and light rays move on
auto-parallel curves which are straight lines. In other words, in any Cartan
gauge $(M,e^{\phi}\eta,\phi)$ there exists a \textquotedblleft
veiled\textquotedblright\ form of special relativity, the natural framework of
which is, of course, Minkowski space-time in the Riemann gauge $(M,\eta,0)$.
Note that a Riemann-Cartan curvature $R{}_{::\beta\mu\nu}^{\alpha}[g,\phi
]\neq0$, that is, a curved Riemann-Cartan space-time, is the necessary and
sufficient condition for existence of a gravitational field.

The second case is given by the class of curved Riemann-Cartan space-times
which are conformally flat in the Riemann gauge, i.e., where $(M,e^{-\phi}%
\eta,0)$ and $R{}_{\:\:\beta\mu\nu}^{\alpha}[e^{-\phi}\eta,0]\neq0$. In such
situations, one can completely gauge away the Riemannian curvature by a Cartan
transformation, thereby going to a Cartan gauge $(M,\eta,\phi)$ with Minkowski
metric$\eta$, which determines the local causal structure, and a torsion
scalar field $\phi$, which generates all dynamical properties of the
gravitational field. This should not be regarded as being equivalent to the
first case above, since the Riemann-Cartan curvature tensor $R{}_{\:\:\beta
\mu\nu}^{\alpha}[\eta,\phi]\neq0$ does not vanish. As in the previous case,
light-rays move along null autoparallel curves that are straight lines, since
the local causal structure is the same as that of Minkowski space-time.
However, test particles move along time-like autoparallel curves that are not
straight lines, considering that in this case the Riemann-Cartan space-time is
curved. This is well illustrated, for instance, when we consider in the
Riemann gauge the class of Robertson-Walker\ (RW) metrics, which are known to
be conformally flat \cite{Ibson}. If we go to the Cartan gauge $(M,\eta,\phi
)$, by means of a Cartan transformation, we arrive at a new cosmological
scenario in which a dynamical metric ceases to determine the cosmic expansion
and other gravitational phenomena, all gravitational effects being now
attributed to torsion through the sole action of a scalar field $\phi$ living
in a curved Riemann-Cartan space-time with a non-dynamical Minkowski metric
$\eta$.

There are many other examples of how distinct physical interpretations of the
same phenomena are possible in different gauges. By way of illustration, we
will consider, in this section, how one would describe, in a general Cartan
gauge, an important effect predicted by general relativity: the so-called
gravitational spectral shift. Let us consider the gravitational field
generated by a massive body, which in an arbitrary Cartan gauge $(M,g,\phi
)$\ is described by both the metric tensor $g_{\mu\nu}$ and the torsion scalar
field $\phi$. For the sake of simplicity, let us restrict ourselves to the
case of a static field, in which neither $g_{\mu\nu}$ nor $\phi$ depends on
time. Let us suppose that a light wave is emitted on the body at a fixed point
with spatial coordinates $(r_{E},\theta_{E},\varphi_{E})$ and received by an
observer at the fixed point $(r_{R},\theta_{R},\varphi_{R}).$ Denoting the
coordinate times of emission and reception by $t_{E}$ and $t_{R}$,
respectively, the light signal, which in the Cartan gauge corresponds to a
null affine geodesic, connects the event $(t_{E},r_{E},\theta_{E},\varphi
_{E})$ with the event $(t_{R},r_{R},\theta_{R},\varphi_{R}).$ Let $\lambda$ be
an affine parameter along this null geodesic with $\lambda=\lambda_{E}$ at the
event of emission and $\lambda=\lambda_{R}$ at the event of reception. If we
write the line element in the form $ds^{2}=g_{00}[r,\theta,\varphi
]dt^{2}-g_{jk}[r,\theta,\varphi]dx^{j}dx^{k}$, then, since the geodesic is
null, we must have
\begin{equation}
g_{00}[r,\theta,\varphi]\left(  \frac{dt}{d\lambda}\right)  ^{2}=
g_{jk}[r,\theta,\varphi]\frac{dx^{j}}{d\lambda}\frac{dx^{k}}{d\lambda},
\label{spectralshift}%
\end{equation}
so we can write
\[
\frac{dt}{d\lambda}=\left(  \frac{g_{jk}[r,\theta,\varphi]}{g_{00}
[r,\theta,\varphi]}\frac{dx^{j}}{d\lambda}\frac{dx^{k}}{d\lambda}\right)
^{\frac{1}{2}}.
\]
On integrating between $\lambda=\lambda_{E}$ and $\lambda=\lambda_{R}$, we
have
\begin{equation}
t_{R}-t_{E}=\int\left(  \frac{g_{jk}[r,\theta,\varphi]}{g_{00} [r,\theta
,\varphi]}\frac{dx^{j}}{d\lambda}\frac{dx^{k}}{d\lambda}\right)  ^{\frac{1}%
{2}}d\lambda\label{spectral2}%
\end{equation}
Because the integral on the right-hand side of the above equation depends only
on the light path through space, and since the emitter and observer are at
fixed positions in space, then $t_{R}-t_{E}$ has the same value for all
signals sent. This implies that for any two signals emitted at coordinate
times $t_{E}^{(1)},\, t_{E}^{(2)}$ and received at $t_{R}^{(1)},\, t_{R}%
^{(2)}$, we have $t_{R}^{(1)}-$ $t_{E}^{(1)}=t_{R}^{(2)}-t_{E}^{(2)}$, which
means that the coordinate time difference $\Delta t_{E}=t_{E}^{(2)}-$
$t_{E}^{(1)}$ at the event of emission is equal to the coordinate time
difference\ $\Delta t_{R}=t_{R}^{(2)}-$ $t_{R}^{(1)}$\ at the event of
reception. On the other hand, we know from Section 3 that the proper time
recorded by clocks in a general Cartan gauge must be calculated by using the
formula
\[
\Delta\tau=\int_{a}^{b}\left(  e^{-\phi}g_{\mu\nu} \frac{dx^{\mu}}{d\lambda
}\frac{dx^{\nu}}{d\lambda}\right)  ^{\frac{1}{2}}d\lambda.
\]
Therefore, the proper time recorded by the clocks of observers situated at the
body and at the point of reception will be given, by the
\[
\Delta\tau_{E}=e^{-\frac{1}{2}\phi_{E}}\sqrt{g_{00}[r_{E},\theta_{E}%
,\varphi_{E}]}\Delta t_{E},
\]
and
\[
\Delta\tau_{R}=e^{-\frac{1}{2}\phi_{R}}\sqrt{g_{00}[r,\theta_{R},\varphi_{R}%
]}\Delta t_{R},
\]
where $\phi_{E}=\phi[r_{E},\theta_{E},\varphi_{E}]$ and $\phi_{R}=\phi
[r_{R},\theta_{R},\varphi_{R}]$. Since $\Delta t_{E}=\Delta t_{R}$, we have
the invariant relation
\[
\frac{\Delta\tau_{R}}{\Delta\tau_{E}}= \frac{e^{-\frac{1}{2}\phi_{R}}%
\sqrt{g_{00}[r_{R},\theta_{R},\varphi_{R}]}}{e^{-\frac{1}{2}\phi_{E}}%
\sqrt{g_{00}[r_{E},\theta_{E},\varphi_{E}]}}.
\]
Suppose now that $n$ waves of frequency $\nu_{E}$ are emitted in proper time
$\Delta\tau_{E}$ from an atom situated on the body. Then $\nu_{E}=\frac
{n}{\Delta\tau_{E}}$ is the proper frequency measured by an observer situated
at the body. On the other hand, the observer situated at the fixed point
$(r_{R},\theta_{R},\varphi_{R})$ will see these $n$ waves in the proper time
$\Delta\tau_{R}$, hence will measure a frequency $\nu_{R}=\frac{n}{\Delta
\tau_{R}}$. Therefore, we have the invariant relation
\begin{equation}
\frac{\nu_{R}}{\nu_{E}}=e^{\frac{1}{2}(\phi_{R}- \phi_{E})}\frac{\sqrt
{g_{00}[r_{E},\theta_{E},\varphi_{E}]}}{\sqrt{g_{00}[r_{R},\theta_{R}%
,\varphi_{R}]}}. \label{spectralshiftweyl}%
\end{equation}

We, thus, see that $\nu_{R}\neq\nu_{E}$, i.e., the observed frequency differs
from the frequency measured at the body, and this constitutes the spectral
shift effect in a general Cartan gauge due to the gravitational field and, in
general, depends on both $g_{00}$ and $\phi$. To conclude, two points related
to the above equation are worth noting. The first is that, since in a Riemann
gauge $\widehat{\phi}=0$, (\ref{spectralshiftweyl}) reduces the well-known
general relativistic formula for the gravitational spectral shift. The second
point is that if we go to a Cartan gauge where $g_{00}$ is constant, then
(\ref{spectralshiftweyl}) becomes simply
\[
\frac{\nu_{R}}{\nu_{E}}=e^{\frac{1}{2}(\phi_{R}-\phi_{E})}.
\]
As we see, in this Cartan gauge all information concerning the gravitational
spectral shift depends only on the torsion scalar field.

\section{Symmetries of space-time and scalar torsion}

In this section, we will consider the notion of space-time symmetry in the
context of Riemann-Cartan space-times endowed with a scalar torsion field,
which, in a certain sense, will generalize some known results of general
relativity concerning isometries and Killing vector fields. We will also
consider the conservation laws of auto-parallel motion in the Riemann-Cartan space-time.

Since in a Cartan gauge the gravitational field is described by two
independent geometric objects, namely, the metric and the scalar torsion
field, the definition of symmetry of the gravitational field should be given
in terms of both. Also, this definition should be invariant under Cartan
transformations and reduce to the usual definition in the Riemann gauge. Thus,
taking into account the action of the Cartan transformations
(\ref{eq:cart trans}) on the fundamental elements $(g_{\mu\nu},\phi)$ of the
gravitational field, we define the symmetries of a Riemann-Cartan space-time
with scalar torsion by the following conditions
\begin{equation}
\mathfrak{L}_{k}g_{\mu\nu}=f\, g_{\mu\nu},\;\mathfrak{L}_{k}\phi=f,
\label{eq:sym def}%
\end{equation}
where $f$ is an arbitrary function and $\mathfrak{L}_{k}$ is the Lie
derivative with respect to the vector field $k^{\mu}=\frac{dx^{\mu}}{d\lambda
}$, referred to as a Killing-Cartan vector field, since it behaves as a
Killing vector field in the Riemann gauge. However, considering that in the
equations (\ref{eq:sym def}) above the Lie derivative of the torsion scalar
field $\mathfrak{L}_{k}\phi=k[\phi]=\phi_{,\mu}k^{\mu}=\frac{d\phi}{d\lambda}$
is equal to the conformal factor $f$ of the metric, the symmetries of a
Riemann-Cartan space-time with scalar torsion can alternatively be defined by
the equations
\begin{equation}
\mathfrak{L}_{k}g_{\mu\nu}=g_{\mu\nu}k[\phi]= g_{\mu\nu}\frac{d\phi}{d\lambda
}, \label{eq:sym def inv}%
\end{equation}
which makes no reference to the arbitrary function $f$ and has the advantage
of being invariant under Cartan transformations. As we have seen, assuming the
invariance of the contravariant components of the Killing-Cartan vector field
under Cartan transformations $\widehat{k}^{\mu}=k^{\mu}=\frac{dx^{\mu}%
}{d\lambda}$ also assures the invariance of the equations
(\ref{eq:sym def inv}) under Cartan transformations (\ref{eq:cart trans}).
Moreover, the definition given by (\ref{eq:sym def inv}) reduces to the
isometry definition in the Riemann gauge, in accordance with the symmetry
definition of the gravitational field in general relativity. Note that the
definition (\ref{eq:sym def inv}) of symmetries of Riemann-Cartan space-times
can also be obtained from the definition $\mathfrak{L}_{k}\eta_{\mu\nu}=0$ of
symmetries of Minkowski space-time in special relativity, through the coupling
prescription $\eta_{\mu\nu}\rightarrow e^{-\phi}g_{\mu\nu}$.

The generalization of the Killing equations in a Cartan gauge can be obtained
from (\ref{eq:sym def inv}) as follows. Substituting the Lie derivative of the
metric, expressed in terms of the Riemann-Cartan covariant derivative, given
by $\mathfrak{L}_{k}g_{\mu\nu}=\widetilde{\nabla}_{\nu}k_{\mu}+ \widetilde
{\nabla}_{\mu}k_{\nu}=k^{\alpha}\nabla_{\alpha}g_{\mu\nu}+ \nabla_{\nu}k_{\mu
}+\nabla_{\mu}k_{\nu}-\frac{1}{2}(\phi_{,\mu}k_{\nu}+\phi_{,\nu}k_{\mu})+
k^{\alpha}\phi_{,\alpha}g_{\mu\nu}$ in (\ref{eq:sym def inv}), and taking into
account the metricity condition $\nabla_{\alpha}g_{\mu\nu}=0$, we obtain that
the Killing-Cartan vector fields must be solution of the equations
\begin{equation}
\nabla_{\mu}k_{\nu}+\nabla_{\nu}k_{\nu}-\frac{1}{2}(k_{\mu}\phi_{,\nu}+
k_{\nu}\phi_{,\mu})=\widetilde{\nabla}_{\nu}k_{\mu}+\widetilde{\nabla}_{\mu
}k_{\nu}- k^{\alpha}\phi_{,\alpha}g_{\mu\nu}=0, \label{eq:kilcar}%
\end{equation}
which will be referred to as Killing-Cartan equations. It is easily seen that
they coincide with the Killing equations in the Riemann gauge where $\phi=0$.
Therefore, the symmetries of the gravitational field in a Cartan gauge are
given by the Killing-Cartan vector fields, which depend on both the metric and
the torsion scalar field. Clearly, the properties of the gravitational field
do not change along the integral curves of these vector fields.

Note that the Killing-Cartan equations (\ref{eq:kilcar}) can also be obtained
from the Killing equations in Minkowski space-time $k_{\nu,\mu}+k_{\mu,\nu}%
=0$, given with respect to a Lorentz coordinate system in special relativity.
Let us consider, by definition, that the contravariant components of the
Killing vector field are invariant under Cartan transformations $\widehat
{k}^{\mu}=k^{\mu}$. Then, from $\widehat{k}_{\nu}=\widehat{g}_{\nu\mu}%
\widehat{k}^{\mu}= e^{-\phi}g_{\nu\mu}k^{\mu}=e^{-\phi}k_{\nu}$ and applying
the coupling prescription (\ref{eq:coupling geo}) to the Killing equations
$k_{\nu,\mu}+k_{\mu,\nu}=0$, we obtain that $\widehat{\nabla}_{\mu}\widehat
{k}_{\nu}+\widehat{\nabla}_{\nu}\widehat{k}_{\mu}= e^{-\phi}[\nabla_{\mu
}k_{\nu}+\nabla_{\nu}k_{\nu} -\frac{1}{2}(k_{\mu}\phi_{,\nu}+k_{\nu}\phi
_{,\mu})]=0$, which are equivalent to the Killing-Cartan equations
(\ref{eq:kilcar}).

It should be noted that the Killing-Cartan vector fields also coincide with
the Killing vector fields in a Cartan gauge where $\mathfrak{L}_{k}\phi
=k[\phi]=\frac{d\phi}{d\lambda}=0$, that is, when the torsion scalar field is
invariant along the integral curve of the Killing-Cartan vector field $k^{\mu
}$, since the Killing-Cartan equations (\ref{eq:sym def inv}) coincide with
the Killing equations.

To conclude this section let us note that the conservation laws of motions
along auto-parallel curves are closely related to the existence of
Killing-Cartan vector fields, and are very useful to solve the equations of
motion of test particles and light rays in a Cartan gauge. Thus, consider a
test particle whose path is an auto-parallel time-like curve $x^{\mu}(\tau)$,
parametrized with proper time $\tau$ and 4-velocity vector $u^{\mu}%
=\frac{dx^{\mu}}{d\tau}$ . Then, the invariant quantity
\begin{equation}
C=e^{-\phi}g_{\mu\nu}k^{\mu}u^{\nu}%
\end{equation}
defined in an arbitrary Cartan gauge, where $k^{\mu}$ is a Killing-Cartan
vector field, is not only an invariant under Cartan transformations, but also
a constant of motion. Considering that%

\begin{equation}
\frac{dC}{d\lambda}=u^{\alpha}\nabla_{\alpha}C=e^{-\phi} [g_{\alpha\beta
}k^{\alpha}u^{\mu}\nabla_{\mu}u^{\beta}- \frac{d\phi}{d\tau}g_{\alpha\beta
}k^{\alpha}u^{\beta}+ \frac{1}{2}(\nabla_{\mu}k_{\nu}+\nabla_{\nu}k_{\nu
})u^{\mu}u^{\nu}]
\end{equation}
and substituting $(\nabla_{\mu}k_{\nu}+\nabla_{\nu}k_{\nu})u^{\mu}u^{\nu}=
\frac{d\phi}{d\tau}g_{\alpha\beta}k^{\alpha}u^{\beta}$ which was obtained from
the Killing-Cartan equations (\ref{eq:kilcar}), we obtain that%

\begin{equation}
\frac{dC}{d\lambda}=u^{\alpha}\nabla_{\alpha}C= e^{-\phi}g_{\alpha\beta
}k^{\alpha}(u^{\mu}\nabla_{\mu}u^{\beta}- \frac{d\phi}{d\tau}u^{\beta})=0,
\end{equation}
taking into account the equations of motion of the test particle given by
(\ref{eq:motion prtime}).

As an example, let us consider, in a Cartan gauge, the gravitational field
equations (\ref{eq: fiedeq ric}) with $T_{\mu\nu}=\Lambda=0$, which is given
by $R_{\mu\nu}[g,\phi]=0$. Clearly, a solution of the above equations is given
by the metric
\begin{equation}
ds^{2}=c^{2}dt^{2}-\frac{1}{(1-\frac{2m}{r})^{2}}dr^{2} -\frac{r^{2}}%
{1-\frac{2m}{r}}(d\theta^{2}+sen^{2}\theta d\psi^{2}) \label{eq: cart sol g}%
\end{equation}
and the torsion scalar field
\begin{equation}
\phi(r)=-ln(1-\frac{2m}{r}). \label{eq:cart sol phi}%
\end{equation}
Note that the solution above is asymptotically flat, since $(1-\frac{2m}%
{r})\rightarrow1$ and $\phi(r)\rightarrow0$ when $r\rightarrow\infty$, while
the constant $m$ can be identified as the mass of the spherical matter
distribution. In the weak field limit ($r\gg2m$), we obtain that the Newtonian
potential (\ref{Newtonian-potential}) of the above solution is given by
$U=-\frac{c^{2}m}{r}$. Since the Newtonian potential of a spherical
distribution of mass with mass $M$ at a distance $r$ from its center is
$U=-\frac{GM}{r}$, where $G$ is the gravitational constant, it follows that
$m=\frac{GM}{c^{2}}$. Surely, this solution is a member of the equivalence
class which corresponds to Schwarschild solution of general relativity. One
can easily verify that the solution (\ref{eq: cart sol g}) and
(\ref{eq:cart sol phi}) can be transformed, through a Cartan transformation
(\ref{eq:cart trans}) with conformal factor $f(r)=ln(1-\frac{2m}{r})$, into
the representation of the gravitational field in the Riemann gauge which is
given by Schwarzschild solution.

This gravitational field has two Killing-Cartan vector fields given by the
time-like vector field $k^{\mu}=\delta_{\:0}^{\mu}$ and the space-like vector
field $w^{\mu}=\delta_{\:3}^{\mu}$, since both $g_{\mu\nu}$ and $\phi$ do not
depend on the coordinates $x^{0}=ct$ and $x^{3}=\psi$. Note that they are also
Killing vector fields, as $k^{\mu}\phi_{,\mu}=0$ and $w^{\mu}\phi_{,\mu}=0$.
Therefore, according to our previous results, a test particle moving in this
gravitational field, along a time-like auto-parallel curve $x^{\mu}(\tau)$
with 4-velocity $u^{\mu}=\frac{dx^{\mu}}{d\tau}$, has the following constants
of motion $E=e^{-\phi}g_{\alpha\beta}k^{\alpha}u{}^{\beta}=(1-\frac{2m}%
{r})u^{0}$ and $L=e^{-\phi}g_{\alpha\beta}w^{\alpha}u^{\beta}=-r^{2}%
sin^{2}\theta u^{3}$ given, respectively, by the energy per unit mass and
angular momentum per unit mass.

The constants of motion $E$ and $L$ above can be used to obtain the equation
of the orbit of a test particle in this gravitational field as follows. Let us
assume, for simplicity, that the particle is moving in the plane $\theta
=\frac{1}{2}\pi$. In a Cartan gauge the norm of the 4-velocity $u^{\mu}%
=\frac{dx^{\mu}}{d\tau}$ of the test particle is such that $g_{\mu\nu}u^{\mu
}u^{\nu}=e^{\phi}$. Thus, using the expressions for $E$ and $L$ above to
substitute the components $u^{0}$=$\frac{dx^{0}}{d\tau}$=$\frac{E}%
{(1-\frac{2m}{r})}$ and $u^{3}=\frac{d\psi}{d\tau}=-\frac{L}{r^{2}}$ into
$g_{\mu\nu}u^{\mu}u^{\nu}=e^{\phi},$ it follows that $E^{2}-(u^{2})^{2}%
-(\frac{L^{2}}{r^{2}}+1)(1-\frac{2m}{r})=0$. Now, substituting $u^{2}%
=\frac{dr}{d\psi}\frac{d\psi}{dr}=-\frac{L}{r^{2}}\frac{dr}{d\psi}$, we obtain that%

\begin{equation}
E^{2}-\frac{L^{2}}{r^{4}}(\frac{dr}{d\psi})^{2}-(\frac{L^{2}}{r^{2}%
}+1)(1-\frac{2m}{r})=0. \label{eq:orbit eq}%
\end{equation}
Finally, integrating the above expression (\ref{eq:orbit eq}) we obtain the
equation of the orbit $r=r(\psi)$, which is identical to the equation of the
orbit in Schwarzschild space-time. Similar results hold with respect to the
motion of light rays. Therefore, the static and spherical symmetric
gravitational field, represented in a Cartan gauge by (\ref{eq: cart sol g})
and (\ref{eq:cart sol phi}), is in accordance with all classical test of
general relativity.

\section{Final remarks}

In this work, we present a scenario in which the gravitational field is not
associated only with the metric tensor, but with the combination of both the
metric $g_{\mu\nu}$ and a geometrical scalar field $\phi$. This formulation
led to a new kind of invariance, which involves simultaneous transformation of
$g_{\mu\nu}$ and $\phi$. Moreover, we have shown that in geometrical setting
the same physical phenomena may appear in different pictures and distinct
representations. For instance, in the case of the gravitational spectral
shift, except in the Riemann gauge, in all other representations the torsion
scalar field plays an essential role. An important conclusion to be drawn from
what has been presented in this paper is that general relativity can be recast
in a Riemann-Cartan space-time. In close connection with this, it should be
mentioned that by using a rather similar procedure it has been shown recently
that general relativity can also be formulated in another non-Riemannian
setting, namely, that of Weyl integrable space-time \cite{Romero}. The
possibilities of enlarging the set of mathematical frameworks that are capable
of describing the same physical theory, in this case, general relativity,
using distinct geometrical languages, seems to give an illustration of the
epistemological view conceived by H. Poincar\'e that the geometry of
space-time is perhaps a convention that can be freely chosen by the
theoretician \cite{Henri,Tavakol}.

\section*{Acknowledgments}

Carlos Romero and Silvina Paola Gomez Martinez would like to thank CNPq and
CAPES for financial support.


\begin{thebibliography}{99}                                                                                               %


\bibitem {Hehl}F.W. Hehl, J.D. McCrea and E.W. Mielke and Y. Ne'eman,
\textit{Phys. Rep}. \textbf{258}, 1 (1995), arxiv:gr-gc/9402012.

\bibitem {Schouten}J.A. Schouten, \textit{Ricci-Calculus} ( Second Edition,
Springer 1954).

\bibitem {Goenner}H. Goenner, \textit{Living Rev. Rel.} \textbf{7}, 2 (2004).

\bibitem {Cartan}E. Cartan, \textit{Comptes Rendus Acad. Sci (Paris)}
\textbf{174}, 437 (1922). For a review with extensive bibliography see F. W.
Hehl, P. von der Heyde, G. D. Kerlick, J. M. Nester, \textit{Rev. Mod. Phys}.
\textbf{48}, 393 (1976). See also M. Gasperini and V. De Sabatta,
\textit{Introduction to Gravitation} (World Scientific, 1986). V. de Sabatta
and C. Sivaram, \textit{Spin and Torsion in Gravitation} (World Scientific,
1994). R. Kerner,\textit{\ Ann. Inst. H. Poincar\'e}, \textbf{34}, 473 (1981).
A. Trautman, \textit{Einstein-Cartan theory} in Encyclopedia of Mathematical
Physics, vol. 2, pages 189--195, Ed. J. P. Fran\c{c}oise, G. L. Naber and Tsou
S.T. (Oxford: Elsevier, 2006). Y. Mao, M. Tegmark, A. H. Guth, and S. Cabi,
Phys. Rev. D76, 104029 (2007), arXiv:gr-qc/0608121.

\bibitem {Aldrovandi}For a review on other gravitational theories making use
of the concept of torsion see R. Aldrovandi and J. G. Pereira, e-Print
arXiv:0801.4148v1. R. Utiyama, \textit{Phys. Rev.} \textbf{101}, 1597 (1956);
T.W.B. Kibble,\textit{\ J. Math. Phys.} 2, 212 (1961); F. W. Hehl, P. Von Der
Heyde, G. D. Kerlick, and J. M. Nester. \textit{Rev. Mod. Phys}., \textbf{4},
393 (1976); A. Trautman, \textit{Symp. Math.} \textbf{12}, 139 (1973); A.
Trautman, arXiv:gr-qc/0606062. F.W. Hehl, \textit{Four lectures on Poincare
gauge field theory}, in Proceedings of the 6th School of Cosmology and
Gravitation, Erice, Italy, (1979). Y.N. Obukhov, Poincar\'e gauge gravity:
Selected topics, Int. J. Geom. Meth. Mod. Phys. 3 (2006) 95-138,
gr-qc/0601090. G. R. Bengochea and R. Ferraro, Phys. Rev. D \textbf{79},
124019 (2009). E. V. Linder, Phys. Rev. D \textbf{81}, 127301 (2010).

\bibitem {fr}For a review, see T. P. Sotiriou and V. Faraoni, Rev. Mod. Phys.
\textbf{82}, 451 (2010) and the references therein.

\bibitem {Capozziello1}S. Capozziello, M. De Laurentis,\textit{ Phys. Rep}.
\textbf{509}, 167 (2011), arXiv:1108.6266.

\bibitem {Mannheim}S. L. Adler, Rev. Mod. Phys. \textbf{54}, 729 (1982). P. D.
Mannheim, Found. Phys. \textbf{42}, 338 (2012). P. D. Mannheim, Gen. Rel.
Grav. \textbf{22}, 289 (1990).

\bibitem {Weyl}H. Weyl, \textit{Sitzungesber Deutsch. Akad. Wiss. Berli} 465
(1918); H. Weyl, \textit{Space, Time, Matter} (Dover, New York, 1952).

\bibitem {Scholz}For a nice review on Weyl geometry see E. Scholz, arXiv:1111.3220.

\bibitem {Novello}M. Novello and H. Heintzmann, \textit{Phys. Lett. A}
\textbf{98}, 10 (1983); K. A. Bronnikov, Yu. M. Konstantinov and V. N.
Melnikov, \textit{Grav. Cosmol.} \textbf{1}, 60 (1995); M. Novello, L.A.R.
Oliveira, J.M. Salim and E. Elbas, \textit{Int. J. Mod. Phys. D} \textbf{1},
641 (1993); J. M. Salim and S. L. Saut\'u, \textit{Class. Quant. Grav}
\textbf{13}, 353 (1996); H. P. de Oliveira, J. M. Salim and S. L. Saut\'u,
\textit{Class. Quant. Grav.} \textbf{14}, 2833 (1997); V. Melnikov,
\textit{Classical Solutions in Multidimensional Cosmology} in
\textit{Proceedings of the VIII Brazilian School of Cosmology and Gravitation
II}, ed. M. Novello (Editions Fronti\`eres, 1995) p.542; R. G. Gannouji, H.
Nandan, N. Dadhich, \textit{JCAP} \textbf{11}, 51 (2011). O. Arias, R.
Cardenas and I.Quiros, \textit{Nucl. Phys. B} \textbf{643}, 187 (2002); J.
Miritzis, \textit{Class. Quant .Grav.} \textbf{21}, 3043 (2004); J. Miritzis,
\textit{J.Phys.: Conf. Ser.} \textbf{8}, 131 (2005); M. Israelit,
\textit{Found. Phys.} \textbf{35}, 1725 (2005); F. Dahia, G. A. T. Gomez and
C. Romero, \textit{J. Math.Phys.} \textbf{49}, 102501 (2008); J. E. Madriz
Aguilar and C. Romero, \textit{Found. Phys.} \textbf{39}, 1205 (2009). T.
Moon, J. Lee, P. Oh, Mod. Phys. Lett. A \textbf{25}, 3129 (2010), arXiv.gr-qc/0912.0432.

\bibitem {Shapiro}G. Germ\`an, Phys. Rev. D \textbf{32}, 3307(1985). I. L.
Shapiro, Phys. Rept.357:113, 2002 and the references therein.

\bibitem {Lambd_transf}A. Einstein, The Meaning of Relativity, Princeton
University Press (1970).

\bibitem {Capozziello2}S. Capozziello, G. Lambiase and C. Stornaiolo,
Geometric classification of the torsion tensor of space-time, arXiv:gr-qc/0101038.

\bibitem {gradtorsion}For a review see Richard T Hammond Rep. Prog. Phys. 65
(2002) 599; Yongsung Yoon, Phys.Rev. D59 (1999) 127501, arXiv:9904018v1; C. J.
Park, Yongsung Yoon, Gen.Rel.Grav. 29 (1997) 765, arXiv: 9611053v1; Sung-Won
Kim Phys. Rev. D 34 (1986), 1011; V. De Sabbata and M. Gasperini Phys. Rev. D
23(1981), 2116; Wei-Tou Ni Phys. Rev. D 19 (1979), 2260; S. Hojman, M.
Rosenbaum, M. P. Ryan, and L. C. Shepley Phys. Rev. D 17 (1978), 3141;

\bibitem {Wald}Robert M. Wald, General Relativity, The University of Chicago
Press (1984).

\bibitem {Rodrigues}J. F. T. Giglio and W. A. Rodrigues Jr., Ann. Phys,
\textbf{524}, 302 (2012). ArXiv: 1111.2206.

\bibitem {Mainwaring}S. R. Mainwaring and G. E. Stedman, \textit{Phys. Rev. A}
\textbf{47}, 3611 (1993).

\bibitem {Poulis}F. P. Poulis and J. M. Salim, \textit{Int. J. Mod. Phys}.:
Conf. Series, \textbf{3}, 87 (2011), arXiv:gr-qc/1106.3031.

\bibitem {BD}C. H. Brans and R. H. Dicke, Phys. Rev. \textbf{124}, 925 (1961).
R. H. Dicke, Phys. Rev. \textbf{125}, 2163 (1962).

\bibitem {Adler}R. Adler, M. Bazin and M. Schiffer, \textit{Introduction to
General Relativity} (McGraw-Hill, 1975).

\bibitem {canuto}Canuto, P. J. Adams, S. H. Hsieh, E. Tsiang, Phys. Rev. D
\textbf{16}, 1643 (1977).

\bibitem {Dabrowski1}M. P. Dabrowski, T. Denkiewicz and D. Blaschke,
\textit{Annalen Phys.} \textbf{16}, 237 (2007)

\bibitem {Faraoni}V. Faraoni, \textit{Cosmology in Scalar-Tensor Gravity}
(Kluwer Academic Publishers, Dordrecht, 2004).

\bibitem {Dirac1}P. A. M. Dirac, \textit{Nature}, \textbf{139}, 323 (1937).

\bibitem {Singh}R. T. Singh and D. Shridhar, \textit{Int. J. Theor. Phys.
}\textbf{26}, 901 (1987).

\bibitem {Pucheu}C. Romero, J. B. Fonseca-Neto and M. L. Pucheu,
\textit{Found. Phys. }\textbf{42},\textit{ }224 (2012), arXiv: gr-qc/1101.5333.

\bibitem {Ibson}M. Ibison, \textit{J. Math. Phys} \textbf{48}, 122501 (2007).

\bibitem {Faraoni2}V. Faraoni, \textit{Int. J. Theor. Phys}, \textbf{38}, 217
(1999). V. Faraoni, E. Gunzig and P. Nardoni, \textit{Fund. Cosm. Phys}.
\textbf{20}, 121 (1999). V. Faraoni and S. Nadeau, \textit{Phys. Rev. D}
\textbf{75}, 23501 (2007).

\bibitem {QuirosAguilar}I. Quiros, R. Bonal and R. Cardenas, Phys. Rev. D
\textbf{62}, 044042 (2000). I. Quiros, R. Garx\'ia-Salcedo and J. E. Madriz
Aguilar, arXiv: gr-qc/1108.2911v2. I. Quiros, R. Garx\'ia-Salcedo, J. E.
Madriz Aguilar and T. Matos, arXiv: gr-qc/1108.5857v2.

\bibitem {HawkingEllis}S.W. Hawking and G.F.R. Ellis, The Large scale
structure of space-time, Cambridge University Press, Cam- bridge (1973).

\bibitem {cassif_torsion}See, for instance, M. Novello, "Theoretical
Cosmology", in \textit{VII Brazilian School of Cosmology and Gravitation, }Ed.
M. Novello (Editions Fronti\`eres, 1995). Iain A. Brown and A. Hammami, arXiv:
gr-qc/1112.0575v2 (2011).

\bibitem {Dabrowski}M. P. Dabrowski, T. Denkiewicz and D. Blaschke,
\textit{Annalen Phys.} \textbf{16}, 237 (2007).

\bibitem {Romero}C. Romero, J. B. Fonseca-Neto and M. L. Pucheu, Class.
Quantum Grav. \textbf{29}, 155015 (2012). ArXiv:1201.1469.

\bibitem {Henri}H. Poincar\'e, \textit{Science and Hypothesis} (Dover, New
York, 1952).

\bibitem {Tavakol}I. W. Roxburgh and R. K. Tavakol, \textit{Found. Phys.}
\textbf{8}, 229 (1978).

\bibitem {Exact Solutions SKMH}H. Stephani, D. Kramer, M.A.H. MacCallum, C.
Hoenselaers and E. Herlt, Exact solutions of Einstein's field equations,
Cambridge Univ. Press, UK (2003).

\bibitem {Kobayashi}S. Kobayashi and K. Nomizu, Foundations of Differential
Geometry, Vol. I, (Interscience Publishers, N.Y., London,1963).
\end{thebibliography}
\end{document}